\documentclass[aps, prx, reprint, superscriptaddress]{revtex4-2}
\pdfoutput=1
\usepackage{braket,dsfont,subfigure,amsfonts,amssymb,bm,graphicx,float,amsmath,amsthm,txfonts}
\usepackage[colorlinks]{hyperref}
\usepackage{orcidlink}
\newcommand{\Tr}{{\rm Tr}}

\newcommand{\rrangle}{\rangle\!\rangle}

\newcommand{\llpipe}{|}
\newcommand{\E}{\mathbb{E}}
\newcommand{\LL}{\mathcal{L}}
\newcommand{\Real}[1]{\ensuremath{\textrm{Re}(#1)}}
\newcommand{\Imag}[1]{\ensuremath{\textrm{Im}(#1)}}
\newcommand{\Z}{\mathbb{Z}}

\newcommand{\sket}[1]{\ensuremath{\llpipe#1\rrangle}}

\theoremstyle{definition}
\newtheorem{definition}{Definition}

\begin{document}
\title{A New Framework for Quantum Phases in Open Systems: Steady State of Imaginary-Time Lindbladian Evolution}
\author{Yuchen Guo~\orcidlink{0000-0002-4901-2737}}
\affiliation{State Key Laboratory of Low Dimensional Quantum Physics and Department of Physics, Tsinghua University, Beijing 100084, China}
\author{Ke Ding~\orcidlink{0009-0002-7788-1224}}
\affiliation{State Key Laboratory of Low Dimensional Quantum Physics and Department of Physics, Tsinghua University, Beijing 100084, China}
\author{Shuo Yang~\orcidlink{0000-0001-9733-8566}}
\email{shuoyang@tsinghua.edu.cn}
\affiliation{State Key Laboratory of Low Dimensional Quantum Physics and Department of Physics, Tsinghua University, Beijing 100084, China}
\affiliation{Frontier Science Center for Quantum Information, Beijing 100084, China}
\affiliation{Hefei National Laboratory, Hefei 230088, China}

\begin{abstract}
    This study delves into the concept of quantum phases in open quantum systems, examining the shortcomings of existing approaches that focus on steady states of Lindbladians and highlighting their limitations in capturing key phase transitions.
    In contrast to these methods, we introduce the concept of imaginary-time Lindbladian evolution as an alternative framework.
    This new approach defines gapped quantum phases in open systems through the spectrum properties of the imaginary-Liouville superoperator.
    We find that, in addition to all pure gapped ground states, the Gibbs state of a stabilizer Hamiltonian at any finite temperature can also be characterized by our scheme, demonstrated through explicit construction.
    Moreover, the closing of the imaginary Liouville gap is associated with the divergence of the Markov length, which has recently been proposed as an indicator of phase transitions in open quantum systems.
    To illustrate the effectiveness of this framework, we apply it to investigate the phase diagram for open systems with $\mathbb{Z}_2^{\sigma}\times \mathbb{Z}_2^{\tau}$ symmetry, including cases with nontrivial average symmetry protected topological order or spontaneous symmetry breaking order.
    Our findings demonstrate universal properties at quantum criticality, such as nonanalytic behaviors of steady-state observables, divergence of correlation lengths, and closing of the imaginary-Liouville gap.
    These results advance our understanding of quantum phase transitions in open quantum systems.
    In contrast, we find that the steady states of real-time Lindbladians do not provide an effective framework for characterizing phase transitions in open systems.
\end{abstract}

\maketitle
\section{Introduction}
Open quantum systems, where the coupling between system and environment cannot be neglected, have attracted much interest in various fields such as condensed matter theory, quantum computing, and quantum information~\cite{Zurek2003, Breuer2007, Horodecki2009, Guehne2009, Rivas2012, Weiss2012, Rotter2015}.
Recent studies focus on extending topologically ordered phases~\cite{Levin2005, Kitaev2006, Chen2010, Zhang2012} in closed systems to open systems, exploring the possibility of long-range entanglement or symmetry protected topological (SPT) phases with specific global symmetries~\cite{Pollmann2010, Chen2011, Schuch2011, Pollmann2012, Levin2012, Chen2013, Chen2014, Senthil2015}.
Significant progress has been made in this direction~\cite{DeGroot2022, Lee2025, Wang2023A, Wang2023B, Ma2023A, Ma2023B, Zhang2023, Hsin2023, Lessa2024A, Ma2024, Wang2024, Guo2024A, Xue2024, Sohal2024, Chirame2024, Ellison2024}, revealing new quantum phenomena and phases that are not observable in closed systems~\cite{Kessler2012, Walter2014, Kessler2021, Lee2023, Lee2025, Lessa2024B, Sala2024, Gu2024, Huang2024, Guo2025, Orito2025, Kuno2025a, Kuno2025b}. 
These investigations could also accelerate the development of advanced quantum computation techniques, including quantum simulation, quantum error mitigation, and error correction~\cite{Preskill2018, Guo2022, Cai2023, Fan2024}.

In closed systems, quantum phases are understood as equivalence classes of gapped quantum states, i.e., the ground states of local, gapped Hamiltonians.
Local unitary evolution is often used as the equivalence relation to define these phases~\cite{Chen2010}.
The correspondence between ground states and Hamiltonians enables the study of phase transitions between different quantum phases by considering a parameterized Hamiltonian $H(g)$ that connects systems belonging to distinct phases.
Physical properties such as the energy gap $\Delta(g)$, correlation length $\xi(g)$, and other observables measured in the corresponding ground states $\ket{\psi(g)}$ can be calculated along this path~\cite{Yang2008, Gu2010, Sachdev2011}.

However, generalizing this framework to open systems, particularly identifying counterparts of these key ingredients, remains an open problem. 
The time evolution of an open system is usually captured by the Lindbladian master equation~\cite{Breuer2007, Rivas2012}.
Some studies propose adopting the steady states of Lindbladians $e^{\int \LL\mathrm{d}t}$ as counterparts to ground states of Hamiltonians, thereby defining dissipative quantum phases and studying possible phase transitions~\cite{Macieszczak2016, Minganti2018, Lieu2020, Liu2024A, Liu2025}.
Meanwhile, the equivalence relation for mixed states is chosen as two-way connectivity by local quantum channels or local Lindbladian evolutions as a generalization of local unitary evolutions~\cite{Sang2024}.
Alternative approaches include using the entanglement gap to replace the energy gap~\cite{Zhou2023}, or studying the stability of a mixed-state phase with the Markov length of conditional mutual information~\cite{Sang2025}.
These methods rely solely on the information of the density matrix itself without requiring access to a `Hamiltonian' or its counterpart in open systems.

In this article, we first demonstrate that the formalism of dissipative quantum phases is insufficient for characterizing many important open-system phase transitions and cannot serve as a general framework in this regime.
Specifically, it fails to degrade to the pure-state definition of quantum phases when returning to a closed system.
Moreover, it is inconsistent with the equivalence relation constructed on the density matrix.
Next, we introduce the concept of imaginary-time Lindbladian evolution $e^{-\int \LL^I\mathrm{d}\tau}$ and formally define the gapped quantum phase in an open system as its steady state with finite recursion time $\tau\sim1/\Delta^I$, where $\Delta^I$ is the spectrum gap of the imaginary-Liouville superoperator $\LL^I$ as the generalization of the energy gap.
We then heuristically construct $\LL^I$ for several typical pure states and mixed states, including the finite-temperature Gibbs states of stabilizer Hamiltonians and quantum states with nontrivial SPT or average SPT (ASPT) order proposed in recent studies~\cite{Ma2023A, Ma2023B}.
We also establish a rigorous relationship between the closing of the imaginary Liouville gap and the divergence of the Markov length in one-dimensional systems with Hermitian jump operators.
Finally, we construct a parameterized model to study the phase diagram and possible phase transitions between different quantum phases protected by $\Z_2^{\sigma}\times \Z_2^{\tau}$ symmetry.
This includes the criticality between the trivial phase, the spontaneous symmetry breaking (SSB) phase, and the ASPT phase.
Remarkably, we identify several properties for quantum criticality in this example, including nonanalytic behaviors of steady-state properties like order parameters and correlation lengths, accompanied by the closing of the imaginary-Liouville gap, validating the effectiveness of our framework for studying phase transitions in open systems.
In contrast, we compare this with the phase diagram obtained from the steady states of real-time Lindbladian evolution.
This reveals the limitations of the real-time approach in capturing phase transitions in open systems, as their steady states are often highly degenerate or can smoothly interpolate between mixed states that essentially belong to distinct phases.

\section{Quantum phases in closed systems}
In closed systems, the time evolution is governed by the Schr\"{o}dinger equation
\begin{align}
    \frac{\mathrm{d}\ket{\psi}}{\mathrm{d}t} = -iH\ket{\psi}.
\end{align}
The quantum phase for a local, gapped Hamiltonian is defined by its ground state properties, which are the zero-temperature equilibrium state
\begin{align}
    \ket{\psi}\hspace{-0.5mm}\bra{\psi} \propto \lim_{\beta \rightarrow \infty}e^{-\beta H}
\end{align}
that can be reached with imaginary-time evolution
\begin{align}
    \ket{\psi}\propto \lim_{\beta \rightarrow \infty}e^{-\beta H}\ket{0}
\end{align}
with a recursion time $\tau\sim 1/\Delta$ and $\Delta$ being the energy gap of $H$.
Therefore, the ground state of a gapped Hamiltonian can be viewed as a steady state of imaginary-time evolution of that Hamiltonian, which constitutes the theory of quantum phase transitions in the following way~\cite{Chen2010, Chen2011}:
\begin{definition}
    Two local, gapped Hamiltonians $H(0)$ and $H(1)$ belong to the same quantum phase iff there exists a smooth path $H(g)$ such that the energy gap $\Delta(g)$ is always finite along the path.
\end{definition}

In other words, a quantum phase transition in a closed system is always accompanied by the occurrence of a gap closing.
This equivalence relation can also be rephrased by the following property of ground states~\cite{Chen2010}.
\begin{definition}
     Two local, gapped Hamiltonians $H(0)$ and $H(1)$ belong to the same quantum phase iff their corresponding ground states $\ket{\psi(0)}$ and $\ket{\psi(1)}$ can be connected by finite-time local evolution $\ket{\psi(1)} =  e^{-i\mathcal{T}\left[\int_0^1\widetilde{H}(g)\mathrm{d}g\right]}\ket{\psi(0)}$ and $\ket{\psi(0)} =  e^{-i\mathcal{T}\left[-\int_0^1\widetilde{H}(g)\mathrm{d}g\right]}\ket{\psi(1)}$.
\end{definition}

\noindent\emph{Remark.} In the presence of ground-state degeneracy, the above definition should be understood as follows.  
For symmetry-breaking phases, one considers the symmetric ground state within the degenerate manifold (e.g., the GHZ state in the transverse-field Ising model at $g=0$).
For topologically ordered phases, the definition applies to any of the topologically degenerate ground states, as none of them can be connected to a trivial product state by finite-time local evolution~\cite{Chen2010, Chen2011}.

\begin{figure*}
    \centering
    \includegraphics[width=0.75\linewidth]{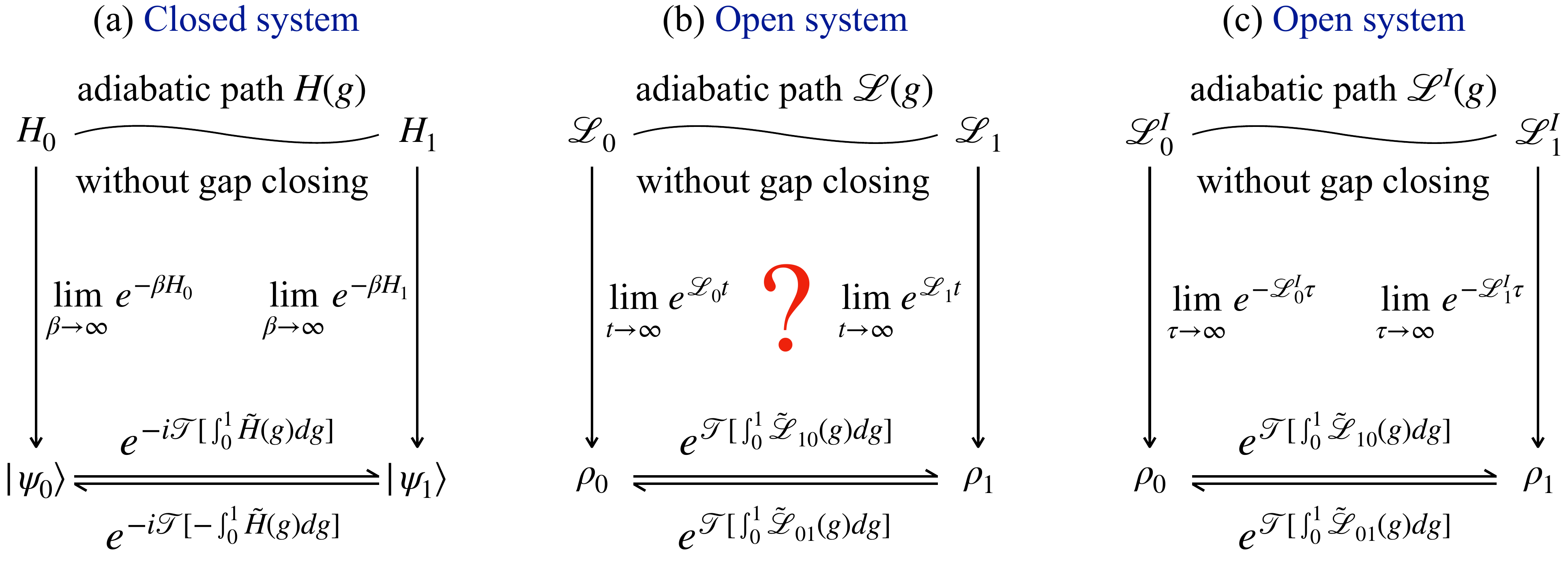}
    \caption{Schematic diagram for the definition of quantum phases in (a) closed systems and (b-c) open systems.}
    \label{Fig: Phase}
\end{figure*}

The equivalence relationship is illustrated in Fig.~\ref{Fig: Phase}(a).
In short, the equivalence between ground states is defined by real-time evolution $ e^{-i\mathcal{T}\left[\int_0^1\widetilde{H}(g)\mathrm{d}g\right]}$, with equivalent Hamiltonians connected by a smooth path without gap closing.
Meanwhile, ground states and Hamiltonians are connected by imaginary-time evolution $\lim_{\beta\rightarrow\infty}e^{-\beta H}$.

\section{Paradox in open systems}
The simplest model to describe the time evolution of a mixed state in an open system is the Lindbladian master equation~\cite{Breuer2007, Rivas2012}
\begin{align}
\begin{aligned}
    \frac{\mathrm{d}\rho}{\mathrm{d}t} &= -i[H, \rho]+\sum_k{\left[L_k\rho L_k^{\dagger}-\frac{1}{2}\{L_k^{\dagger}L_k, \rho\}\right]}\\
    &= -iH_{\rm eff}\rho+i\rho H_{\rm eff}^{\dagger}+\sum_k{L_k\rho L_k^{\dagger}} \equiv \LL(\rho),\label{Equ: Lindblad}
\end{aligned}
\end{align}
where $H$ is the Hamiltonian of the system, and $L_k$ is the jump operator to describe the coupling between system and environment.
In some cases, a non-Hermitian Hamiltonian $H_{\rm eff} = H-\sum_k\frac{i}{2}L_k^{\dagger}L_k$ is introduced to construct an effective theory of the original open system to simplify the analysis.
However, this will not be pursued further in this paper.
These components constitute the Liouville map $\LL$ that acts on the linear space of density matrices.
In particular, the locality condition is imposed on both $H$ and $L_k$.

We now introduce the notion of the supervector for a mixed state.
A general mixed state can be expressed as an eigenvalue decomposition 
\begin{align}
    \rho = \sum_k{\lambda_k\ket{\psi_k}\hspace{-1mm}\bra{\psi_k}},
\end{align}
which naturally corresponds to an unnormalized supervector
\begin{align}
    \sket{\rho} = \sum_k{\lambda_k\ket{\psi_k}\otimes\ket{\psi_k^*}}.
\end{align}
With this formalism, one can rewrite the above linear map of a density matrix $\LL(\rho)$ as a superoperator acting on a supervector $\LL\sket{\rho}$~\cite{Shang2024}
\begin{align}
    \LL\sket{\rho} = \left(-iH_{\rm eff}\otimes I + iI\otimes H_{\rm eff}^{*}+\sum_k L_k \otimes L_k^{*}\right)\sket{\rho}.\label{Equ: Liouville}
\end{align}
Both forms of the Liouville $\LL$ will be utilized in the rest of this paper.
In this section, we briefly review several recent attempts towards a complete paradigm for the open-system quantum phases.

\subsection{Dissipative quantum phases}
Generally, the Liouville superoperator $\LL$ is a non-Hermitian operator whose spectrum is rather complicated.
Suppose that it can be diagonalized with the following form to define eigenvalues and (right) eigenstates
\begin{align}
    \LL\sket{\rho_i} = \lambda_i\sket{\rho_i}.
\end{align}
It has been proven that~\cite{Minganti2018}
\begin{itemize}
    \item $\Real{\lambda_i}\leq 0$.
    \item $\Imag{\lambda_i} = 0$ if $\rho_i$ is Hermitian.
\end{itemize}
The dissipative quantum phase is defined for the steady state of a local, gapped (defined as follows) Lindbladian evolution in real time, i.e., $\rho_{\rm ss} = \lim_{t\rightarrow +\infty} e^{\LL t}\rho(0)$ for a general initial state $\rho(0)$ not orthogonal to $\rho_{\rm ss}$, as shown in Fig.~\ref{Fig: Phase}(b).
From the above properties, we know that $\rho_{\rm ss}$ should be the eigenvector of $\LL$ satisfying $\LL\sket{\rho_{\rm ss}} = 0$ (if it exists).
We can rearrange these eigenvalues according to their real parts as $0=\lambda_0\geq \Real{\lambda_1}\geq \Real{\lambda_2}\cdots$ and the Liouville gap is defined as $\Delta = -\Real{\lambda_1}$.
As a consequence, a finite Liouville gap corresponds to a steady state of the Lindbladian evolution achieved with recursion time $T\sim 1/\Delta$.

At first glance, this definition is similar to that in closed systems, and there have been many previous studies that considered a parametrized $\LL(g)$ and calculated the corresponding steady state $\rho(g)$ to study possible phase transitions~\cite{Minganti2018}.
However, we note that it does not contain the definition of pure states. 
If we take $L_k=0$, the Lindbladian equation will degrade to the Schr\"{o}dinger equation, where all the eigenstates of $H$ become the steady state of $e^\LL \mathrm{d}t$ with zero eigenvalues.
The underlying reason is that such a dissipative quantum phase is defined by the steady state of a real-time evolution instead of imaginary-time evolution, which is not naturally consistent with the pure-state case and will also lead to other unexpected consequences, as discussed below.

\subsection{Equivalence relation for mixed states}
Recently, it has been proposed to utilize finite-time local Lindbladian evolution as the equivalence relation for the classification of quantum phases in mixed states~\cite{Sang2024, Sang2025}.
Specifically, if two mixed states $\rho(0)$ and $\rho(1)$ belong to the same quantum phase, they can be bidirectionally connected via finite-time local Lindbladian evolution
\begin{align}
    \sket{\rho(1)} = e^{\mathcal{T}\left[\int_{0}^{1} \widetilde{\LL}_{10}(g)\mathrm{d}g\right]}\sket{\rho(0)},\\    \sket{\rho(0)} = e^{\mathcal{T}\left[\int_{0}^{1} \widetilde{\LL}_{01}(g)\mathrm{d}g\right]}\sket{\rho(1)}.
\end{align}
From this definition, it is clear that all the mixed states with a finite Liouville gap belong to the same equivalence class as they can be generated by finite-time Lindbladian evolutions from each other.
To show the paradox, consider two open systems $\LL(0)$ and $\LL(1)$, each having a finite Liouville gap.
Their steady states are denoted $\rho(0)$ and $\rho(1)$, respectively.
Although the Liouville gap may close along a path $\LL(g)$ connecting $\LL(0)$ and $\LL(1)$, it is still possible to establish bidirectional connections between $\rho(0)$ and $\rho(1)$ via finite-time Lindbladian evolutions by setting $\widetilde{\LL}_{01} = \LL(0)$ and $\widetilde{\LL}_{10} = \LL(1)$.
This is because $\rho(i)$ can be achieved through $\LL(i)$ from any initial state not orthogonal to it for $i=0, 1$.
This implies that the Liouville gap cannot be used to define distinct phases in open systems, as all states can be connected to a product state and therefore belong to the same trivial phase.

In summary, though real-time Lindbladian evolution describes realistic dissipative dynamics, it does not provide a direct description for the phase transitions in mixed states because a) it cannot degrade to the pure-state case, and b) it is not consistent with the definition of the equivalence class of mixed states via real-time Lindbladian evolution.

\section{Quantum phases in open systems: imaginary-time Lindbladian evolution}
As discussed earlier, we cannot use the same real-time Lindbladian evolution to define both quantum phases and the equivalence relation for mixed states.
The experience of closed systems inspires us to consider imaginary-time evolution to define quantum phases in open systems~\cite{Kamakari2022, Khindanov2024}.

\subsection{Imaginary-time Lindbladian evolution}
Here, we try to derive the imaginary-time evolution for an open system, which cannot be directly achieved by applying the wick rotation to the Lindbladian equation in Eq.~\eqref{Equ: Lindblad} due to the lack of unitarity.
The key idea is to consider the imaginary-time evolution for the entire system (system and environment) and trace out the ancillary degrees of freedom.
An effective model to simulate the (real-time) Lindbladian equation for a general multipartite system is multipartite collision models~\cite{Cattaneo2021, Cattaneo2023}, which we utilize to derive the imaginary-time form of the Lindbladian equation.

For each jump operator $L_k$, we introduce a two-level ancillary qubit initiated in the state $\ket{0}\hspace{-0.5mm}\bra{0}$.
For simplicity, we focus on the case with only one $L_k$ and consider the following real-time evolution for time step $\Delta t$~\cite{Cattaneo2021, Cattaneo2023}
\begin{align}
    \begin{aligned}
        \rho(t+\Delta t)_{\rm s} = & \,\Tr_{\rm a}\left[e^{-iH_{\rm s}\otimes I_{\rm a} \Delta t - iL_{k \rm s}\otimes \sigma^{+}_{\rm a}\sqrt{\Delta t} - iL_{k \rm s}^{\dagger}\otimes \sigma^{-}_{\rm a}\sqrt{\Delta t}}\right.\\
        &\,\rho(t)_{\rm s}\otimes \ket{0}\hspace{-0.5mm}\bra{0}_{\rm a}
        \left.e^{iH_{\rm s}\otimes I_{\rm a} \Delta t + iL_{k \rm s}^{\dagger}\otimes \sigma^{-}_{\rm a}\sqrt{\Delta t} + iL_{k \rm s}\otimes \sigma^{+}_{\rm a}\sqrt{\Delta t}}\right]\\
        = &\,\Tr_{\rm a}\left[\left(\rho(t)_{\rm s}-iH_{\rm s}\rho(t)_{\rm s}\Delta t+i\rho(t)_{\rm s}H_{\rm s}\Delta t\right)\otimes\ket{0}\hspace{-0.5mm}\bra{0}_{\rm a}\right.\\
        - & \,iL_{k\rm s}\rho(t)_{\rm s}\otimes \ket{1}\hspace{-0.5mm}\bra{0}_{\rm a}\sqrt{\Delta t}+i\rho(t)_{\rm s}L_{k\rm s}^{\dagger}\otimes \ket{0}\hspace{-0.5mm}\bra{1}_{\rm a}\sqrt{\Delta t}\\
        + & \,L_{k\rm s}\rho(t)_{\rm s}L_{k\rm s}^{\dagger}\otimes \ket{1}\hspace{-0.5mm}\bra{1}_{\rm a}\Delta t\\
        - & \left.\frac{1}{2}\left(L_{k\rm s}^{\dagger}L_{k\rm s}\rho(t)_{\rm s}+ \rho(t)_{\rm s}L_{k\rm s}^{\dagger}L_{k\rm s}\right)\otimes \ket{0}\hspace{-0.5mm}\bra{0}_{\rm a}\Delta t\right]\\
        = & \,\rho(t)_{\rm s}-i\left[H_{\rm s}, \rho(t)_{\rm s}\right]\Delta t\\
        + & \,\sum_k\left[ L_{k\rm s}\rho(t)_{\rm s}L_{k\rm s}^{\dagger}-\frac{1}{2}\left\{L_{k\rm s}^{\dagger}L_{k\rm s}, \rho(t)_{\rm s}\right\}\right]\Delta t,
    \end{aligned}
\end{align}
which is just the Lindbladian equation in Eq.~\eqref{Equ: Lindblad}.

Similarly, we can apply the same technique to derive the imaginary-time Lindbladian evolution as follows
\begin{align}
    \begin{aligned}
        \rho(\tau+\Delta \tau)_{\rm s} = & \,\Tr_{\rm a}\left[e^{-H_{\rm s}\otimes I_{\rm a} \Delta \tau - L_{k \rm s}\otimes \sigma^{+}_{\rm a}\sqrt{\Delta \tau} - L_{k \rm s}^{\dagger}\otimes \sigma^{-}_{\rm a}\sqrt{\Delta \tau}}\right.\\
        &\,\rho(\tau)_{\rm s}\otimes \ket{0}\hspace{-0.5mm}\bra{0}_{\rm a}
        \left.e^{-H_{\rm s}\otimes I_{\rm a} \Delta \tau - L_{k \rm s}^{\dagger}\otimes \sigma^{-}_{\rm a}\sqrt{\Delta \tau} - L_{k \rm s}\otimes \sigma^{+}_{\rm a}\sqrt{\Delta \tau}}\right]\\
        = &\,\Tr_{\rm a}\left[\left(\rho(\tau)_{\rm s}-H_{\rm s}\rho(\tau)_{\rm s}\Delta \tau-\rho(\tau)_{\rm s}H_{\rm s}\Delta \tau\right)\otimes\ket{0}\hspace{-0.5mm}\bra{0}_{\rm a}\right.\\
        - & \,L_{k\rm s}\rho(\tau)_{\rm s}\otimes \ket{1}\hspace{-0.5mm}\bra{0}_{\rm a}\sqrt{\Delta \tau}-\rho(\tau)_{\rm s}L_{k\rm s}^{\dagger}\otimes \ket{0}\hspace{-0.5mm}\bra{1}_{\rm a}\sqrt{\Delta \tau}\\
        + & \,L_{k\rm s}\rho(\tau)_{\rm s}L_{k\rm s}^{\dagger}\otimes \ket{1}\hspace{-0.5mm}\bra{1}_{\rm a}\Delta \tau\\
        + & \left.\frac{1}{2}\left(L_{k\rm s}^{\dagger}L_{k\rm s}\rho(\tau)_{\rm s}+ \rho(\tau)_{\rm s}L_{k\rm s}^{\dagger}L_{k\rm s}\right)\otimes \ket{0}\hspace{-0.5mm}\bra{0}_{\rm a}\Delta \tau\right]\\
        = & \,\rho(\tau)_{\rm s}-\left\{H_{\rm s}, \rho(\tau)_{\rm s}\right\}\Delta \tau\\
        + & \,\sum_k\left[ L_{k\rm s}\rho(\tau)_{\rm s}L_{k\rm s}^{\dagger}+\frac{1}{2}\left\{L_{k\rm s}^{\dagger}L_{k\rm s}, \rho(\tau)_{\rm s}\right\}\right]\Delta \tau,
    \end{aligned}
\end{align}
from which we obtain the imaginary-time Lindbladian evolution as
\begin{align}
\begin{aligned}
    \frac{\mathrm{d}\rho}{\mathrm{d}\tau} &= -\left\{H, \rho\right\}+ \sum_k \left[L_k \rho L_k^{\dagger} + \frac{1}{2}\{L_k^{\dagger}L_k, \rho\}\right]\\
    &=-\left\{H_{\rm eff}^I, \rho\right\} + \sum_k L_k \rho L_k^{\dagger} \equiv - \LL^I(\rho),\label{Equ: imag}
\end{aligned}
\end{align}
or in the superoperator form
\begin{align}
    \LL^I\sket{\rho} = \left(H_{\rm eff}^I\otimes I + I\otimes H_{\rm eff}^{I*}-\sum_k L_k \otimes L_k^{*}\right)\sket{\rho},
\end{align}
where $H_{\rm eff}^I = H - \sum_k \frac{1}{2}L_k^{\dagger}L_k$ becomes a Hermitian operator in this case, and we refer to $\LL^I$ as the imaginary-Liouville superoperator.
The above differential equation corresponds to the following integral form
\begin{align}
    \sket{\rho(\tau)} = e^{-\int_0^{\tau}\LL^I\mathrm{d}\tau}\sket{\rho(0)}.
\end{align}
It is obvious that the mapping $e^{-\LL^{I}\mathrm{d}\tau}$ is Hermitian preserving.
In addition, it is completely positive given that the imaginary-time evolution of the entire system is completely positive.
Notably, it does not directly satisfy the trace-preserving condition and requires renormalization after each step of evolution, similar to the imaginary-time evolution of a Hamiltonian in a closed system.
Therefore, if it evolved from a valid density matrix, the steady state of imaginary-time evolution is also a legal density matrix that satisfies the Hermitian and positive conditions.

\subsection{Spectrum analysis of imaginary-Liouville superoperator}
An important problem arises immediately: whether one can directly solve the spectrum of $\LL^I$ to identify the steady state of the imaginary-time evolution.
Similar to the study for real-time Lindbladian evolution~\cite{Macieszczak2016, Minganti2018}, we will prove that the (right) eigenvector of the following equation
\begin{align}
    \LL^I{(\rho)} = E\rho, \quad \textrm{or} \quad  \LL^I\sket{\rho} = E\sket{\rho}\label{Equ: Eigen}
\end{align}
satisfies the Hermitian condition for a physical density matrix if the eigenvalue is real, or can be linearly recombined to satisfy this condition in the degenerate case.
However, this argument does not hold for a general complex eigenvalue.

To begin with, we note that in supervector form, the Hermitian condition for the density matrix can be interpreted as a symmetry constraint that permuting the ket and bra of $\rho$ (referred to as $\mathcal{S}$) should keep the density matrix invariant or formally written as $\mathcal{S}\sket{\rho}=\sket{\rho}$.
This symmetry superoperator is antiunitary with $\mathcal{S}^2=\mathcal{I}$, whose eigenvalue is a complex number with unit magnitude $e^{i\theta}$, where $\theta$ is denoted by convention as a symmetry charge.
The critical issue here is the irrelevance of a specific symmetry charge since the supervector possesses an arbitrary global phase.
For example, if one encounters a supervector with $\mathcal{S}\sket{\rho}=e^{i\theta}\sket{\rho}$, then we can redefine $\sket{\rho^{\prime}} = e^{i\frac{\theta}{2}}\sket{\rho}$ to eliminate the phase and lead to a Hermitian density matrix $\rho^{\prime}$.
In other words, the only thing that matters is to check that the supervector is symmetric under the symmetry action $\mathcal{S}$ regardless of its charge $\theta$.

Next, we can verify that this symmetry superoperator commutes with our imaginary-Liouville superoperator as $S\LL^I=\LL^IS$.
With the standard analysis procedure
\begin{align}
    \LL^I\mathcal{S}\sket{\rho}=\mathcal{S}\LL^I\sket{\rho} = \mathcal{S}E\sket{\rho} = E^*\mathcal{S}\sket{\rho},
\end{align}
we know that $\mathcal{S}\sket{\rho}$ is also an eigenvector of $\LL^I$ with eigenvalue $E^*$.
It leads to several possibilities:
\begin{enumerate}
    \item $E$ is a real number and the eigenvector $\sket{\rho}$ is non-degenerate, hence $\mathcal{S}\sket{\rho}\propto\sket{\rho}$.
    \item $E$ is a real number but degenerate, then the degenerate eigenvectors span an invariant subspace of $\mathcal{S}$.
    \item $E$ is a complex number with nonzero imaginary part, which means that $\mathcal{S}\sket{\rho}=\sket{\rho^{\dagger}}$ is also an eigenvector of $\LL^I$ with eigenvalue $E^*$.
\end{enumerate}
In the first case, one directly comes to the conclusion that $\rho$ must preserve the Hermitian condition in the density matrix form.
In the second case with degeneracy, we may consider a linear combination of eigenstates to prevent the occurrence of symmetry breaking.
However, if $E$ is a general complex number as in the third case, we only know that the eigenvalues appear in conjugate pairs, but we cannot discuss any property of the corresponding eigenvectors.
On the other hand, the Hermiticity of an eigenvector also promises that the corresponding eigenvalue will be a real number, which can be demonstrated by taking the Hermitian conjugation of Eq.~\eqref{Equ: Eigen}.
Therefore, we can conclude that
\begin{align}
    E\in \mathbb{R} \Longleftrightarrow \rho = \rho^{\dagger},
\end{align}
where linear combination may be required to obtain a Hermitian $\rho$ in the degenerate case.

However, both eigenstates with real and complex $E$ are involved in the imaginary-time evolution.
To illustrate the physical implications, we begin by assuming that $\mathcal{L}^I$ is diagonalizable.
Although the right eigenvectors $\rho_i$ of a general non-Hermitian superoperator $\mathcal{L}^I$ are not necessarily orthogonal, they remain linearly independent and therefore span the operator space.
Consequently, the initial density matrix $\rho(0)$ can be expressed as a linear combination of the $\rho_i$, i.e.,
\begin{align}
    \rho(0) = \sum_i a_i\rho_i + \sum_j \left(b_j\rho_j + b_j^{*}\rho_j^{\dagger}\right),
\end{align}
where $\rho_i$ are eigenstates with real $E_i$, while $\rho_j$ and $\rho_j^{\dagger}$ are eigenstates with the conjugate pairs $E_j$ and $E_j^*$, respectively.
The imaginary-time Lindbladian evolution of the above state is given by
\begin{align}
\begin{aligned}
    \rho(\tau) &\propto \sum_ie^{-E_i\tau}a_i \rho_i \\
    &+ \sum_j e^{-\Real{E_j}\tau}\left(e^{-i\Imag{E_j}\tau}b_j\rho_j + e^{i\Imag{E_j}\tau}b_j^{*}\rho_j^{\dagger}\right).
\end{aligned}
\end{align}
Each term has a damping rate proportional to the real part of $E_i$, regardless of whether it is real or complex.

On the other hand, due to its non-Hermiticity, $\mathcal{L}^I$ may not be diagonalizable, as is the case at exceptional points (EPs)—a prominent topic in non-Hermitian physics that has attracted considerable attention~\cite{Heiss2012, Miri2019, Oezdemir2019}.
At an EP, the geometric multiplicity (i.e., the number of linearly independent eigenvectors corresponding to one eigenvalue~\cite{Nering1970}) is strictly less than the algebraic multiplicity (i.e., the degeneracy of this eigenvalue in terms of the characteristic polynomial of a matrix), implying that the right eigenvectors of $\mathcal{L}^I$ span only a subspace of the full operator space~\cite{Zhang2019}.
As a result, if the initial state lies outside this subspace, generalized eigenvectors must be included to complete the basis.
Consequently, the time evolution acquires algebraic corrections to the exponential form, which can substantially affect the convergence dynamics.
While convergence to a steady state may still occur, its approach can be qualitatively modified compared to the diagonalizable case.

In summary, away from EP, one can diagonalize $\LL^I$ and sort the spectrum according to its real part as $\Real{E_0}\leq \Real{E_1}\leq \Real{E_2}\leq\cdots$.
As a result, the imaginary-time Lindbladian evolution $-e^{\LL^I\mathrm{d}\tau}$ has a steady state if and only if $E_0$ is a real number.
Otherwise, the relative phase between $e^{-i\Imag{E_j}\tau}b_j\rho_j$ and $e^{i\Imag{E_j}\tau}b_j^{*}\rho_j^{\dagger}$ will never converge for $\tau\rightarrow \infty$.
Meanwhile, since all eigenstates of $\LL^I$ are included in the time evolution as discussed above, we should define the imaginary-Liouville gap as $\Delta^I = \Real{E_1}-E_0$, with the recursion time for the imaginary-time evolution $\tau^I\sim 1/\Delta^I$.
In particular, this definition naturally degrades to the case of closed systems by taking $L_k=0$, where the imaginary-Liouville gap just equals the conventional energy gap, which justifies its rationality.
This formalism also allows us to define `excited states' for an open system according to the real part of the $\LL^I$ spectrum.
Furthermore, complete positivity of $e^{-\LL^I \mathrm{d}\tau}$ guarantees that steady state $\rho_0$ satisfies the positivity condition.
However, we note that a general argument about positivity cannot be made for an excited state even if it has a real eigenvalue $E$.

\subsection{Special case: equilibrium Gibbs states of stabilizer Hamiltonians}
An important feature of our imaginary-time formalism is the natural emergence of a statistical interpretation.
Specifically, $H_{\rm eff}^I$ spans the energy levels and $L_k$ induces quantum jumps between different eigenstates of $H_{\rm eff}^I$ to achieve a statistical equilibrium.
Therefore, we expect that our formalism is suitable for characterizing thermal equilibrium states at any finite temperature, the major class of mixed states in real-world scenarios.
Here, the inverse temperature $\beta$ is determined by the strength of quantum jumps $L_k$.
In this section, we provide an explicit construction of the imaginary-time Lindbladian evolution $e^{-\LL^I\mathrm{d}\tau}$, whose steady state is the equilibrium Gibbs state for a given stabilizer Hamiltonian $H$ at inverse temperature $\beta$ as $\rho\propto e^{-\beta H}$.

A stabilizer Hamiltonian for an $N$-qubit system is defined as $H=\sum_{i}^{N} h_i$, which is composed of $N$ linearly independent operators belonging to the Pauli group $h_i \in G_N$ that commute with each other.
Therefore, the eigenstate of $H$ is the common eigenstate of each $h_i$ with a specific charge $s_i=\pm 1$.
In this sense, the set of $\{s_i\}$ serves as a unique label for an eigenstate $\ket{\psi_{\{s_i\}}}$.
If we refer to $s_i=1$ as an elementary excitation, then the total energy for the state $\ket{\psi_{\{s_i\}}}$ can be easily represented as
\begin{align}
    E_{\{s_i\}} = \sum_i^N s_i = -N+2\alpha,
\end{align}
where $\alpha$ is the number of excitations.
The corresponding Gibbs state at temperature $\beta$ is
\begin{align}
    \rho = \frac{1}{Z}\sum_{\{s_i\}}e^{-\beta E_{\{s_i\}}}\ket{\psi_{\{s_i\}}}\hspace{-0.5mm}\bra{\psi_{\{s_i\}}},\label{Equ: Gibbs}
\end{align}
where $Z=\sum_{\{s_i\}}e^{-\beta E_{\{s_i\}}}$ is the partition function.

An interesting property of a stabilizer group is the existence of an excitation operator $o_i\in G_N$ for each term $h_i$ satisfying that $o_ih_j = (-1)^{\delta_{ij}}h_j o_i$, i.e., $o_i$ only anti-commutes with $h_i$ but commutes with other $h_j$ for $j\neq i$.
In other words, $o_i$ is capable of connecting different eigenstates that differ only by one charge $s_i$.
Inspired by the statistical understanding of our imaginary-Liouville $\LL^I$, we just take $H$ to be the original stabilizer Hamiltonian and take $L_i=\sqrt{\gamma} o_i$ as each excitation operator with constant $\gamma$ to describe the coupling strength.
In this case, we have
\begin{align}
    H_{\rm eff}^I = \sum_i^N h_i - \frac{N}{2}\gamma\sim \sum_i^N h_i,
\end{align}
where the constant term is neglected.
Consider the implementation of $\LL^I$ on the above Gibbs state
\begin{align}
\begin{aligned}
    \LL^I\left(\rho\right)&= \frac{1}{Z}\sum_{\{s_i\}}e^{-\beta E_{\{s_i\}}} \\
    & \left( 2E_{\{s_i\}}\ket{\psi_{\{s_i\}}}\hspace{-0.5mm}\bra{\psi_{\{s_i\}}} -\gamma \sum_{j=1}^{N}o_j\ket{\psi_{\{s_i\}}}\hspace{-0.5mm}\bra{\psi_{\{s_i\}}}o_j^{\dagger} \right)\\
    & \equiv \sum_{\{s_i\}} c_i \ket{\psi_{\{s_i\}}}\hspace{-0.5mm}\bra{\psi_{\{s_i\}}},
\end{aligned}
\end{align}
in which the coefficient of a specific eigenstate $\ket{\psi_{\{s_i\}}}$ is denoted as $c_i$.
It can be calculated as
\begin{align}
    c_i = \frac{1}{Z}\left(e^{-\beta E_{\{s_i\}}} 2E_{\{s_i\}}-\sum_{j=1}^Ne^{-\beta E_{\{s_i^{\prime}(j)\}}} \gamma\right),
\end{align}
where $\{s_i^{\prime}(j)\}$ describes the state related to $\{s_i\}$ by $o_j$.
If $\ket{\psi_{\{s_i\}}}$ has $\alpha$ excitations, then there are $\alpha$ states in the set of $\ket{\psi_{\{s_i^{\prime}(j)\}}}$ with $(\alpha-1)$ excitations and $(N-\alpha)$ states with $(\alpha+1)$ excitations.
Consequently, the coefficient of $\ket{\psi_{\{s_i\}}}$ after applying $\LL^I$ becomes
\begin{align}
    \begin{aligned}
    c_i &= \frac{e^{-\beta E_{\{s_i\}}}}{Z}\left[2(-N+2\alpha)-\alpha \gamma e^{2\beta}-(N-\alpha) \gamma e^{-2\beta}\right]\\
    &= \frac{e^{-\beta E_{\{s_i\}}}}{Z}\left\{\alpha[4-2\sinh{(2\beta)}\gamma]-2N-e^{-2\beta}N\gamma\right\}.
    \end{aligned}
\end{align}
Therefore, $\rho$ is an eigenstate of $\LL^I$ if the coupling strength satisfies that
\begin{align}
    \gamma = \frac{2}{\sinh{(2\beta)}} \label{Equ: Coupling}
\end{align}
with the eigenvalue being
\begin{align}
    E = 2N-e^{-2\beta}N\gamma = -N\sqrt{\gamma^2+4},\label{Equ: GSE}
\end{align}.

On the other hand, we note that if we define
\begin{align}
    \mathcal{L}^I_i = h_i\otimes I +I\otimes h_i^{*} - \gamma o_i\otimes o_i^{*},
\end{align}
we have $\left[\mathcal{L}^I_i, \mathcal{L}^I_j\right]=0$, meaning that they share the same eigenvector.
In addition, direct diagonalization of each $\LL^{I}_i$ gives a spectrum of $\varepsilon=\left\{\pm \sqrt{\gamma^2+4}, \pm \gamma\right\}$.
Comparison with Eq.~\eqref{Equ: GSE} leads to the conclusion that the Gibbs state in Eq.~\eqref{Equ: Gibbs} is indeed the steady state of the imaginary-time Lindbladian evolution constructed here.
In particular, since the derivation of both the real-time and imaginary-time Lindbladian formalisms relies on the weak-coupling assumption, the correspondence described above is exact in the low-temperature regime and becomes only approximate as the temperature increases.
Nevertheless, this example demonstrates the capacity of our formalism to characterize both ground states and finite-temperature equilibrium states, paralleling the way real-time Lindbladian evolution gives access to thermal ensembles under detailed balance conditions~\cite{Palmero2019, Chen2023, Sun2024, Andrew2025}.

\subsection{Quantum phases in open systems}
With the definition of imaginary-time Lindbladian evolution and analysis of its basic properties, we can formally define the quantum phase for open systems:
\begin{definition}
    Two local, gapped open systems described by $\LL^I(0)$ and $\LL^I(1)$ belong to the same quantum phase iff there exists a smooth path $\LL^I(g)$ such that the imaginary-Liouville gap $\Delta^I(g)$ is always finite along the path.
\end{definition}
We now conjecture that this definition is equivalent to another one directly defined in the mixed state~\cite{Sang2024, Sang2025}.
\begin{definition}
    Two local, gapped open systems described by $\LL^I(0)$ and $\LL^I(1)$ belong to the same quantum phase iff the corresponding steady states $\rho(0)$ and $\rho(1)$ can be two-way connected by local Lindbladian evolutions of finite time $\sket{\rho(0)} = e^{\mathcal{T}\left[\int_{0}^{1} \widetilde{\LL}_{01}(g)\mathrm{d}g\right]}\sket{\rho(1)}$ and $
    \sket{\rho(1)} = e^{\mathcal{T}\left[\int_{0}^{1} \widetilde{\LL}_{10}(g)\mathrm{d}g\right]}\sket{\rho(0)}$.
\end{definition}

In summary, our formalism is represented in Fig.~\ref{Fig: Phase}(c), which we believe is the correct generalization of Fig.~\ref{Fig: Phase}(a) into the regime of open systems instead of Fig.~\ref{Fig: Phase}(b).
In the following section, we demonstrate that this conjecture can be validated under specific conditions.
Furthermore, our subsequent numerical simulations, along with the results reported in Ref.~\cite{Guo2025}, provide additional support for this conjecture in several concrete models.

\subsection{Gapless imaginary-Liouville superoperator and divergent Markov length}
\begin{figure*}
    \centering
    \includegraphics[width=0.8\linewidth]{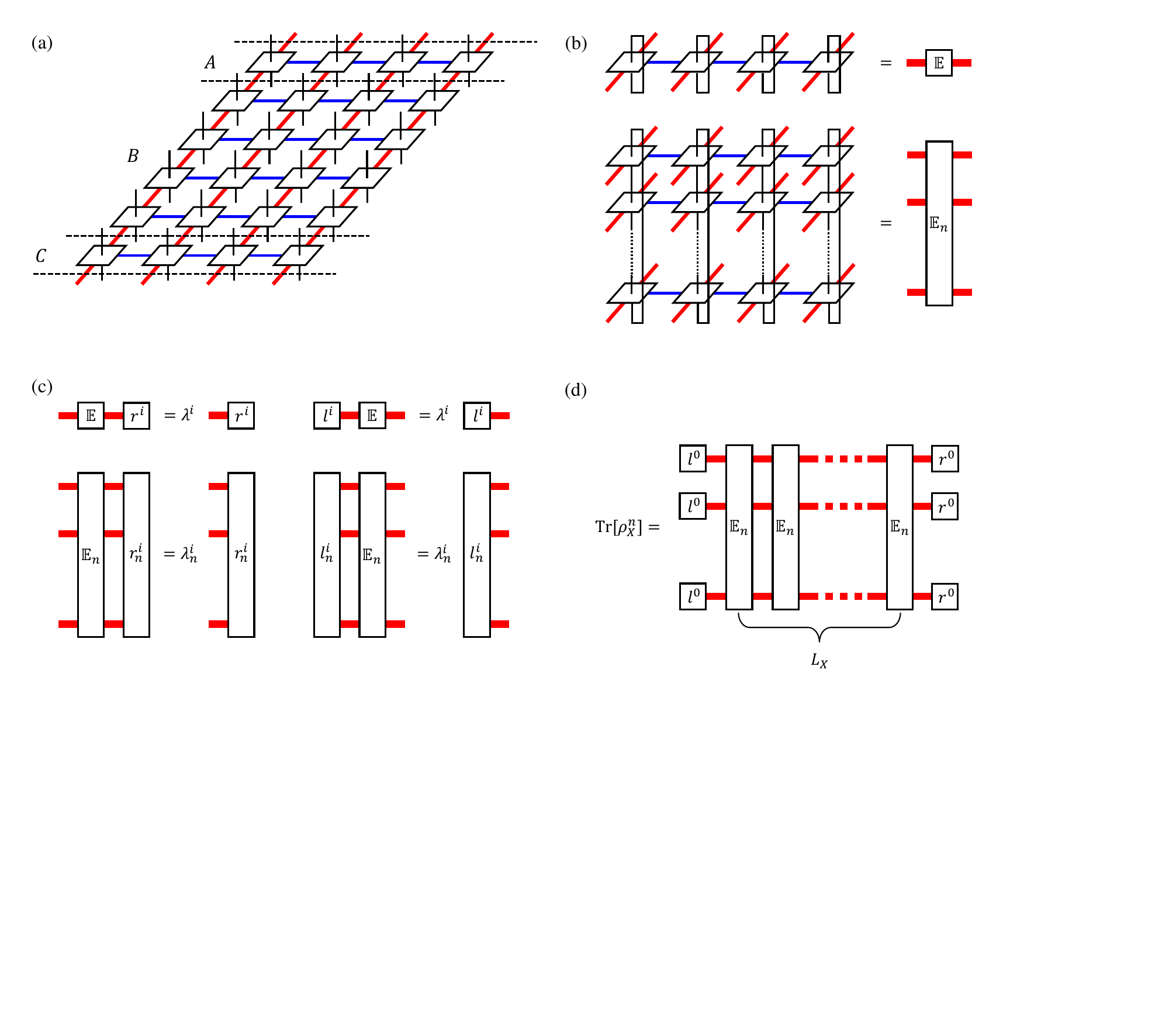}
    \caption{Calculation of CMI and Markov length for a mixed state represented by a uniform PEPO in 2D open systems.
    (a) PEPO representation and division of subregions.
    (b) 1D Transfer operator $\E$ and $\E_n$ for linear and R\'enyi-$n$ correlators (even $n$).
    We have grouped the virtual indices (red legs) of different sites, represented by thicker legs.
    (c) Eigenvalues and left or right eigenvectors for $\E$ and $\E_n$.
    (d) Calculation of R\'enyi-$n$ entropy with tensor contraction.}
    \label{Fig: Markov}
\end{figure*}
A recent study~\cite{Sang2025} proposed the Markov length $\xi_{\rm M}$ as an indicator of quantum phase transitions in open systems, which is defined as the correlation length associated with the conditional mutual information (CMI) $I(A:C|B)$ between subregions $A$ and $C$ that are completely separated by $B$.
Specifically, starting from a short-range correlated mixed state evolved under a local Lindbladian, it was shown that the existence of a quasi-local reverse Lindbladian is guaranteed if and only if the Markov length $\xi_{\rm M}$ remains finite throughout the evolution, implying that the system remains within the same quantum phase. 
In this work, we prove that for one-dimensional (1D) systems with Hermitian jump operators $L_k = L_k^{\dagger}$, a vanishing imaginary Liouville gap $\Delta^I$ necessarily leads to a divergence in the Markov length, establishing a rigorous connection between gap closing and long-range correlations.

We begin by discussing the properties of the Markov length and its relation to conventional correlation lengths in general spatial dimensions.
CMI for three subregions is defined as
\begin{align}
    I(A:C|B) = S(AB)+S(BC)-S(B)-S(ABC),
\end{align}
where $S(X)=-\Tr\left[\rho_X\ln \rho_X\right]$ refers to the von Neumann entropy of the reduced density matrix $\rho_X$.
A representative partition is illustrated in Fig.~\ref{Fig: Markov}(a) for a two-dimensional (2D) system, where Markov length $\xi_{\rm M}$ characterizes the long-range decay of CMI with respect to the distance between subregions $A$ and $C$, or the width of the subregion $B$, i.e.,
\begin{align}
    I(A:C|B)\sim e^{-L_B/\xi_{\rm M}}.
\end{align}
However, evaluating the von Neumann entropy requires full knowledge of the spectrum of reduced density matrices, which presents substantial challenges for both analytical and numerical treatments.
To address this, we consider an alternative definition based on the R\'enyi-$n$ entropy $S^{(n)}(X) = \frac{1}{1-n}\ln\Tr\left[\rho_X^n\right]$, which leads to the corresponding R\'enyi CMI and Markov length~\cite{Kuno2025}
\begin{align}
\begin{aligned}
    I^{(n)}(A:C|B) &= S^{(n)}(AB)+S^{(n)}(BC)-S^{(n)}(B)-S^{(n)}(ABC)\\
    &\sim e^{-L_B/\xi_{\rm M}^{(n)}},
\end{aligned}
\end{align}
where the extrapolation $n\rightarrow 1$ recovers the original von Neumann version of $I$ and $\xi_{\rm M}$.

A finite Markov length serves as an indicator of a ``gapped'' quantum phase in open systems, and thus can be regarded as a mixed-state analog of the conventional correlation length.
Nevertheless, various forms of correlation functions and their associated correlation lengths can still be defined for mixed states.
To proceed, we consider two types of observables.
The linear observable is defined through the standard expectation value
\begin{align}
    \braket{O}\equiv \Tr[\rho O],
\end{align}
while the R\'enyi-$n$ observable is defined as~\cite{Liu2024B}
\begin{align}
    \braket{O}^{(n)}\equiv \frac{\Tr{\left[\rho^{\frac{n}{2}}O\rho^{\frac{n}{2}}O\right]}}{\Tr\left[\rho^n\right]}.
\end{align}
In particular, the R\'enyi-$2$ observable corresponds to the expectation value of the supervector $\sket{\rho}$~\cite{Guo2024A, Lessa2024B}, while the R\'enyi-$1$ observable, also known as the Wightman observable, is defined under canonical purification~\cite{Liu2024B, Weinstein2024}.
Subsequently, their corresponding correlators can be constructed as follows, including the linear correlation function
\begin{align}
\mathcal{C}(O, i, j) \equiv \braket{O_i O_j} - \braket{O_i}\braket{O_j}\sim \ e^{-|i-j|/\xi},\label{Linear_corr}
\end{align}
and the R\'enyi-$n$ correlation function
\begin{align}
    \mathcal{C}^{(n)}(O, i, j)\equiv \braket{O_i O_j}^{(n)} - \braket{O_i}^{(n)}\braket{O_j}^{(n)}\sim\ e^{-|i-j|/\xi_{n}},
\label{Renyin}
\end{align}
which define the correlation lengths $\xi$ and $\xi_n$, respectively.

A natural question thus arises: what is the relationship between the set of R\'enyi-$n$ Markov lengths $\xi_{\rm M}^{(n)}$ and the set of R\'enyi-$n$ correlation lengths $\xi_n$?
We show that, at least for mixed states well approximated by tensor network representations, these two types of correlation lengths coincide for even values of $n$.
For a uniform tensor network, such as the projected entangled pair operator (PEPO) in 2D systems shown in Fig.~\ref{Fig: Markov}(a), the R\'enyi-$n$ correlation length $\xi_n$ is determined by the spectral properties of the associated one-dimensional transfer operator $\E_n$ in Fig.~\ref{Fig: Markov}(b) for even $n$. 
For odd $n$, the operator $\rho^{n/2}$ in the correlator generally cannot be represented by a PEPO with finite bond dimension, hence the argument does not hold.
Specifically, suppose that the spectral decomposition of $\E_n$ reads $\E_n=\sum_{i}\lambda_n^i|{r_n^i})({l_n^i}|$, where the eigenvalues are ordered in decreasing magnitude $|\lambda_n^0|\geq |\lambda_n^1|\geq \cdots$, as shown in Fig.~\ref{Fig: Markov}(c), then the R\'enyi-$n$ correlation length is determined by $\xi_n = -1/ \ln |\lambda_n^1/\lambda_n^0|$.
For the calculation of CMI, we note that $\Tr\left[\rho_X^n\right]$ can be represented as the tensor contraction in Fig.~\ref{Fig: Markov}(d), or formally written as
\begin{align}
\begin{aligned}
    \Tr\left[\rho_X^n\right] &= (l^{0\,{\otimes n}}| E_n^{L_X}|r^{0\,{\otimes n}})\\
    &=\sum_i(\lambda_n^i)^{L_X}  (l^{0\,{\otimes n}}|{r_n^i})({l_n^i}|r^{0\,{\otimes n}})\equiv \sum_i(\lambda_n^i)^{L_X} f_{n}^i,
\end{aligned}
\end{align}
where we denote $f_{n}^i\equiv (l^{0\,{\otimes n}}|{r_n^i})({l_n^i}|r^{0\,{\otimes n}})$.
Therefore, $S^{(n)}(X)$ can be expanded as
\begin{align}
\begin{aligned}
    S^{(n)}(X) &= \frac{1}{1-n} \ln \Tr\left[\rho_X^n\right]\\
    &\approx \frac{1}{1-n}\ln \left\{(\lambda_n^0)^{L_X}f_n^0\left[1 + \left(\frac{\lambda_n^1}{\lambda_n^0}\right)^{L_X}\frac{f_n^1}{f_n^0}\right]\right\}\\
    &\approx \frac{1}{1-n}\left[L_X \ln \lambda_n^0+\ln f_n^0+\frac{f_n^1}{f_n^0}e^{-L_X/\xi_n}\right].
\end{aligned}
\end{align}
Here, in the first approximation, we only preserve the two largest eigenvalues of $\E_n$ in terms of magnitude, while we adopt $\ln(1+x)\approx x$ in the second.
Finally, we obtain the expression of CMI defined by R\'enyi-$n$ entropy
\begin{align}
\begin{aligned}
    &\,I^{(n)}(A:C|B)\\
    =&\,\frac{1}{1-n}\frac{f_n^1}{f_n^0}\left[e^{-L_{AB}/\xi_n}+e^{-L_{BC}/\xi_n}-e^{-L_{B}/\xi_n}-e^{-L_{ABC}/\xi_n}\right]\\
    =&\,\frac{1}{n-1}\frac{f_n^1}{f_n^0}\left(1-e^{-L_{A}/\xi_n}\right)\left(1-e^{-L_{C}/\xi_n}\right)e^{-L_{B}/\xi_n}\\
    \propto & \,e^{-L_{B}/\xi_n},
\end{aligned}
\end{align}
from which we immediately reach the conclusion that $\xi_{\rm M}^{(n)}=\xi_n$ for even $n$.

Thus far, our discussion has been carried out in arbitrary spatial dimensions, where Rényi correlators of different orders may exhibit distinct asymptotic behaviors~\cite{Lessa2024B}, rendering a general extrapolation to the limit $n\rightarrow 1$ intractable.
However, in 1D systems, a stronger structure emerges.
Specifically, Ref.~\cite{Guo2024A} established that in 1D systems, a strongly injective mixed state, i.e., one with a finite R\'enyi-$2$ correlation length $\xi_2$, necessarily satisfies weak injectivity, implying a finite linear correlation length $\xi$.
In other words, if a mixed state exhibits a divergent $\xi_{n}$, i.e., $\xi$ of $\rho^n$, then $\xi_{2n}$ ($\xi_2$ of $\rho^n$) also diverges.
Motivated by this structure, we now restrict our analysis to 1D systems.
In this setting, we show that a vanishing imaginary Liouville gap—under the condition that the jump operators $L_k$ are Hermitian—implies that the corresponding double state exhibits a divergent correlation length, as established in Ref.~\cite{Hastings2006}.
This correlation length corresponds precisely to the R\'enyi-$2$ correlation length $\xi_2$ of the density matrix.
From this, it follows that higher-order R\'enyi correlation lengths such as $\xi_4, \xi_8, \cdots$ must also diverge, finally leading to a divergent Markov length $\xi_{\rm M}$ by extrapolating to $n\rightarrow 1$, thus validating our conjecture.
In summary, under certain conditions, we justify the proposed correspondence between the spectral properties of the imaginary Liouville superoperator and the equivalence relation to define quantum phases in open systems.

We note that the above correspondence between the closing of the imaginary-Liouville gap and the divergence of the Markov length is established using the well-defined hierarchy of different correlation lengths in 1D systems~\cite{Guo2024A}, such that a divergent Rényi-$2$ correlation length necessarily leads to a divergence of all even-$n$ Rényi correlation lengths and of the $n\rightarrow 1$ correlation length.
This implication may fail in higher dimensions with intrinsic complexity.
Even in closed systems, a projected entangled pair state (PEPS) with finite bond dimension does not always correspond to a short-range correlated or gapped state~\cite{Verstraete2006, Perez2008, Wahl2013, Yang2015}.
Therefore, the present proof should be considered rigorous only in 1D, while its generalization to higher dimensions remains an open question.

As for the relation between these two criteria in a more general case, we can also consider the physical picture behind our imaginary-time Lindbladian evolution to illustrate.
Regarding the derivation of our formalism, we consider the imaginary-time evolution of the entire system and then trace out the environmental degrees of freedom, mimicking the standard procedure of real-time Lindbladian evolution.
In the weak-coupling limit, an intuitive analogy between the imaginary-time Lindbladian steady state and the ground state of the total Hamiltonian can be established.
However, since the derivation involves reinitialization of ancillary degrees of freedom, this correspondence should not be taken literally.
The physical picture nevertheless provides an intuition for why closing the imaginary-Liouville gap signals a critical point in the purified-state description, accompanied by divergent correlation lengths.

\section{Example: Phase transition for ASPT phases}
Recently, studies have constructed a fixed point mixed state for a nontrivial topological phase in an open system by tuning part of a quantum superposition into a classical mixture~\cite{Ma2023B, Guo2024A}.
Here, we demonstrate that such a mixed state can be easily realized as a steady state of our imaginary-time Lindbladian evolution.

\subsection{Decohered Haldane phase}
In Ref.~\cite{Guo2024A}, the authors proposed a fixed-point tensor construction for the so-called ASPT phase protected by $\Z_2^{\sigma}\times \Z_2^{\tau}$ symmetry in a $(1+1)D$ open system.
We start from the cluster state in the closed system, which possesses nontrivial SPT order under the same symmetry (the well-known Haldane phase), whose Hamiltonian is
\begin{align}
    H=-\sum_i\left(\sigma_i^z \tau_{i+1 / 2}^x\sigma_{i+1}^z +\tau_{i-1 / 2}^z \sigma_i^x \tau_{i+1 / 2}^z\right).
\end{align}
In this model, there are two spin-$\frac{1}{2}$ degrees of freedom $\sigma_i$ and $\tau_{i+1/2}$ at the site $i$, each carrying a linear representation of one of $\Z_2$.
In particular, $U = \prod_i \sigma_i^x$ and $K = \prod_i \tau_{i+1/2}^x$ define two global $\Z_2$ symmetries of the system.

The ground state of this model follows the decorated domain wall construction~\cite{Chen2014},
\begin{align}
\begin{aligned}
    \ket{\psi^{\rm SPT}} &= \frac{1}{2^{N/2}}\sum_{\{\sigma_i\}}\ket{\cdots\uparrow_\sigma\rightarrow_{\tau}\uparrow_{\sigma}\rightarrow_{\tau}\uparrow_{\sigma}\leftarrow_{\tau}\downarrow_{\sigma}\leftarrow_{\tau}\uparrow_{\sigma}\cdots}\\
    &\equiv \frac{1}{2^{N/2}}\sum_{\{\sigma_i\}}\ket{\psi_{\{\sigma_i\}}^{\rm SPT}},
\end{aligned}\label{Equ: Cluster}
\end{align}
where excitations of $\tau$ spins are placed at the domain wall of $\sigma$ spins.
Specifically, all components $\ket{\psi_{\{\sigma_i\}}^{\rm SPT}}$ span the degenerate ground state manifold of $H_1\equiv-\sum_i\sigma_i^z\tau_{i+1/2}^x\sigma_{i+1}^z$, while $H_2 \equiv \sum_i\tau_{i-1 / 2}^z \sigma_i^x \tau_{i+1 / 2}^z$ introduces a map between different components (a type of local deforming rule), which determine the superposition coefficients in the final ground state of $H$.

If one of the $\Z_2$ symmetries is broken to weak symmetry due to decoherence, the system still exhibits nontrivial topological properties jointly protected by strong and weak symmetries.
The corresponding density matrix can be constructed as 
\begin{align}
    \rho^{\rm ASPT} =  \frac{1}{2^{N}}\sum_{\{\sigma_i\}}\ket{\psi_{\{\sigma_i\}}^{\rm SPT}}\hspace{-0.5mm}\bra{\psi_{\{\sigma_i\}}^{\rm SPT}},\label{Equ: DM}
\end{align}
where $K = \prod_i \tau_{i+1/2}^x$ remains a strong symmetry, defined as $K\rho=e^{i\theta}\rho$, while $U = \prod_i \sigma_i^x$ becomes a weak symmetry only satisfying that $U\rho U^{\dagger}=\rho$~\cite{Ma2023A}.
It is denoted as the decohered Haldane phase, since it has the same symmetry group structure and symmetry actions on the boundary as the conventional Haldane phase in closed systems~\cite{Guo2024A}.

\subsection{Construction of imaginary-Liouville superoperator}
To construct the imaginary-time Lindbladian evolution that realizes this mixed state as its steady state, we utilize the observation that the classical mixture in the density matrix form can also be viewed as a type of superposition in the supervector space.
In the following, we discuss a heuristic method to construct $H$ and $L_k$ for such a density matrix.

\subsubsection{Single-qubit symmetric state}
We begin by considering a single-qubit system with $H=-\sigma^x$, whose ground state is $\ket{\psi}=\ket{\rightarrow}$.
From an alternative perspective, $H_1=0$ defines a degenerate ground-state manifold $\{\ket{\uparrow}, \ket{\downarrow}\}$, while $H_2 = -\sigma^x$ induces mapping within this manifold and determines the superposition coefficients of different components.
As a result, we obtain the ground state of $H$, namely $\ket{\psi}=\frac{1}{\sqrt{2}}\left(\ket{\uparrow}+\ket{\downarrow}\right)$.

On the other hand, what happens if one tries to construct the open system to realize the classical mixture, i.e., $\rho = \frac{1}{2}\left(\ket{\uparrow}\hspace{-0.5mm}\bra{\uparrow}+\ket{\downarrow}\hspace{-0.5mm}\bra{\downarrow}\right)$?
From the experience with pure states, we expect one term to project out the ground-state manifold spanned by $\{\ket{\uparrow}\hspace{-0.5mm}\bra{\uparrow}, \ket{\downarrow}\hspace{-0.5mm}\bra{\downarrow}\}$, while the other term determines the mixture coefficients.
Therefore, we choose $H=0$, $L^{[1]}=\sigma^z$, and $L^{[2]}=\sigma^x$ to realize this construction, where $L^{[1]}=\sigma^z$ forces the bra and ket spins to align.
It can be verified that the unique steady state of $e^{-\LL^I\mathrm{d}\tau}$ as defined in Eq.~\eqref{Equ: imag} is indeed the given state $\rho = \frac{1}{2}\left(\ket{\uparrow}\hspace{-0.5mm}\bra{\uparrow}+\ket{\downarrow}\hspace{-0.5mm}\bra{\downarrow}\right)$.

\subsubsection{Symmetric product state}
To proceed, we will try to construct an open system for a decohered symmetric product state.
Specifically, we consider the corresponding trivial state with the same symmetry action as that of Eq.~\eqref{Equ: Cluster} (pure state) or Eq.~\eqref{Equ: DM} (mixed state).
The symmetric product state with $\Z_2^{\sigma}\times \Z_2^{\tau}$ symmetry is just
\begin{align}
    \ket{\psi^{\rm trivial}} = \ket{\cdots\rightarrow_{\sigma}\rightarrow_{\tau}\rightarrow_{\sigma}\rightarrow_{\tau}\rightarrow_{\sigma}\rightarrow_{\tau}\cdots}\label{Equ: Trivial}
\end{align}
or written in a consistent form with Eq.~\eqref{Equ: Cluster} as
\begin{align}
    \begin{aligned}
       \ket{\psi^{\rm trivial}} &= \frac{1}{2^{N/2}}\sum_{\{\sigma_i\}}\ket{\cdots\uparrow_\sigma\rightarrow_{\tau}\uparrow_{\sigma}\rightarrow_{\tau}\uparrow_{\sigma}\rightarrow_{\tau}\downarrow_{\sigma}\rightarrow_{\tau}\uparrow_{\sigma}\cdots}\\
    &\equiv \frac{1}{2^{N/2}}\sum_{\{\sigma_i\}}\ket{\psi_{\{\sigma_i\}}^{\rm trivial}}, 
    \end{aligned}
\end{align}
whose Hamiltonian is given by
\begin{align}
    H = -\sum_i {\tau}_{i+1/2}^x-\sum_i\sigma_i^x.
\end{align}
It can be interpreted as $H_1=-\sum_i {\tau}_{i+1/2}^x$ spanning the degenerate ground-state manifold $\{\ket{\psi_{\{\sigma_i\}}^{\rm trivial}}\}$, while $H_2=-\sum_i \sigma_i^x$ inducing mapping between different terms.
Decohering $\sigma$ spins in the above pure state will result in 
\begin{align}
\begin{aligned}
    \rho^{\rm trivial} &=  \frac{1}{2^{N}}\sum_{\{\sigma_i\}}\ket{\psi_{\{\sigma_i\}}^{\rm trivial}}\hspace{-0.5mm}\bra{\psi_{\{\sigma_i\}}^{\rm trivial}}\\
    &=\cdots\frac{1}{2}\left(\ket{\uparrow}\hspace{-0.5mm}\bra{\uparrow}+\ket{\downarrow}\hspace{-0.5mm}\bra{\downarrow}\right)_{\sigma}\otimes\ket{\rightarrow}\hspace{-0.5mm}\bra{\rightarrow}_{\tau}\cdots,
\end{aligned}
\end{align}
with a strong symmetry $K = \prod_i \tau_{i+1/2}^x$ and a weak symmetry $U = \prod_i \sigma_i^x$, i.e., a mixed state with the same symmetry action as that of Eq.~\eqref{Equ: DM} but with trivial topological properties.

The first step to construct the corresponding imaginary-Liouville is to stabilize the ground-state manifold spanned by $\{\ket{\psi_{\{\sigma_i\}}^{\rm trivial}}\hspace{-0.5mm}\bra{\psi_{\{\sigma_i\}}^{\rm trivial}}\}$ for every $\{\sigma_i\}$ configuration, which can be achieved by
\begin{align}
     H &= - \sum_{i}\tau_{i+1/2}^x, \quad L_{i}^{[1]} = \sigma_i^z,
\end{align}
where we have added a superscript $[1]$ since there might be more than one jump operator located on each site.
On the other hand, the hopping terms between different components within the ground-state manifold are chosen as
\begin{align}
    L_{i}^{[2]} = \sigma_i^x.
\end{align}
It is noted that $L_k^{\dagger}L_k=I$ in this case, which means that the term $\sum_k\frac{1}{2}L_k^{\dagger}L_k$ only contributes to a global energy shift to $H_{\rm eff}^{I}$.
We would like to emphasize that now there are in total $4N$ terms (where $N$ is the number of sites) in $\LL^I$ and they commute with each other, capable of stabilizing a $4N$-qubit supervector (a $2N$-qubit density matrix with $N$ sites and $2$ spins at each site).
Moreover, the difference between strong and weak symmetries can be explicitly manifested in this formalism.
By definition, $K$ must commute with $H_{\rm eff}$ and each of $L_k$, while $U$ only needs to commute with $H_{\rm eff}$, with an additional phase allowed for $UL_k = e^{i\phi_k}L_kU$.

\subsubsection{Decohered ASPT state}
Now we are ready to prepare the open system for the density matrix in Eq.~\eqref{Equ: DM} belonging to the decohered Haldane phase.
Similar to the above case, the stabilizers for the ground-state manifold are
\begin{align}
     H &= - \sum_{i}\sigma_i^z\tau_{i+1/2}^x\sigma_{i+1}^z, \quad L_{i}^{[1]} = \sigma_i^z,
\end{align}
while the hopping terms between different components are given by
\begin{align}
    L_{i}^{[2]} = \tau_{i-1/2}^z\sigma_i^x\tau_{i+1/2}^z, 
\end{align}
where $L_k^{\dagger}L_k=I$ is also satisfied in this case, together with the symmetry properties
\begin{align}
    &[K, H] = 0,\quad [K, L_k] = 0,\label{Equ: Strong}\\
    &[U, H] = 0, \quad [U, L_{i}^{[2]}] = 0, \quad UL_{i}^{[1]} = -L_{i}^{[1]}U\label{Equ: Weak}.
\end{align}

\subsection{Phase diagram}
In summary, we have considered the following four imaginary Livioulles $\LL^I$ in an open system with $\Z_2^{\sigma}\times \Z_2^{\tau}$ symmetry.
\begin{align}
    \textrm{Trivial pure: } & H^{00} = -\sum_{i}\tau_{i+1/2}^x-\sum_{i}\sigma_{i}^x, \nonumber \\
    & L^{00} = 0 ,\label{Equ: HL1}\\
    \textrm{SPT: } & H^{01} = - \sum_{i}\sigma_i^z\tau_{i+1/2}^x\sigma_{i+1}^z - \sum_{i}\tau_{i-1/2}^z\sigma_i^x\tau_{i+1/2}^z , \nonumber \\
    & L^{01}=0 ,\\
    \textrm{Trivial mixed: } & H^{10} = - \sum_{i}\tau_{i+1/2}^x, \nonumber \\
    & L^{[1]10}_{i} = \sigma_i^z, \ L^{[2]10}_{i} = \sigma_i^x ,\\
    \textrm{ASPT: } & H^{11} = - \sum_{i}\sigma_i^z\tau_{i+1/2}^x\sigma_{i+1}^z, \nonumber\\
    & L^{[1]11}_{i} = \sigma_i^z, \ L^{[2]11}_{i} = \tau_{i-1/2}^z\sigma_i^x\tau_{i+1/2}^z.
    \label{Equ: HL2}
\end{align}
We linearly interpolate between these four systems
\begin{align}
\begin{aligned}
    \LL^I(\alpha, \beta) &= (1-\alpha)(1-\beta)\LL^I_{00}+(1-\alpha)\beta \LL^I_{01} \\
    & +\alpha(1-\beta)\LL^I_{10}+\alpha\beta\LL^I_{11},
\end{aligned}
\end{align}
with the corresponding Hamiltonian and jump operators becoming
\begin{align}
\begin{aligned}
    H(\alpha, \beta) = (1-\beta)\left(-\sum_{i}\tau_{i+1/2}^x\right)+\beta\left(-\sum_{i}\sigma_i^z\tau_{i+1/2}^x\sigma_{i+1}^z\right)\\
    +(1-\alpha)\left[(1-\beta)\left(-\sum_{i}\sigma_{i}^x\right)+\beta\left(- \sum_{i}\tau_{i-1/2}^z\sigma_i^x\tau_{i+1/2}^z\right)\right]\label{Equ: Ham}
\end{aligned}
\end{align}
and
\begin{align}
\begin{aligned}
    L_{i}^{[1]}(\alpha, \beta)&=\sqrt{\alpha}\sigma_i^z ,\\
    L_{i}^{[2]}(\alpha, \beta)&=\sqrt{\alpha(1-\beta)}\sigma_i^x ,\\
    L_{i}^{[3]}(\alpha, \beta)&=\sqrt{\alpha\beta}\tau_{i-1/2}^z\sigma_i^x\tau_{i+1/2}^z.\label{Equ: Jump}
\end{aligned}
\end{align}
Here, $\alpha\in [0, 1]$ and $\beta\in [0, 1]$ constitute a two-dimensional parameter space, where the path along $\alpha=0$ reveals the conventional phase transition between the Haldane phase and the trivial phase, while the system has a mixed steady state for any $\alpha>0$.

\subsection{Numerical simulations}
\begin{figure*}
    \includegraphics[width=\linewidth]{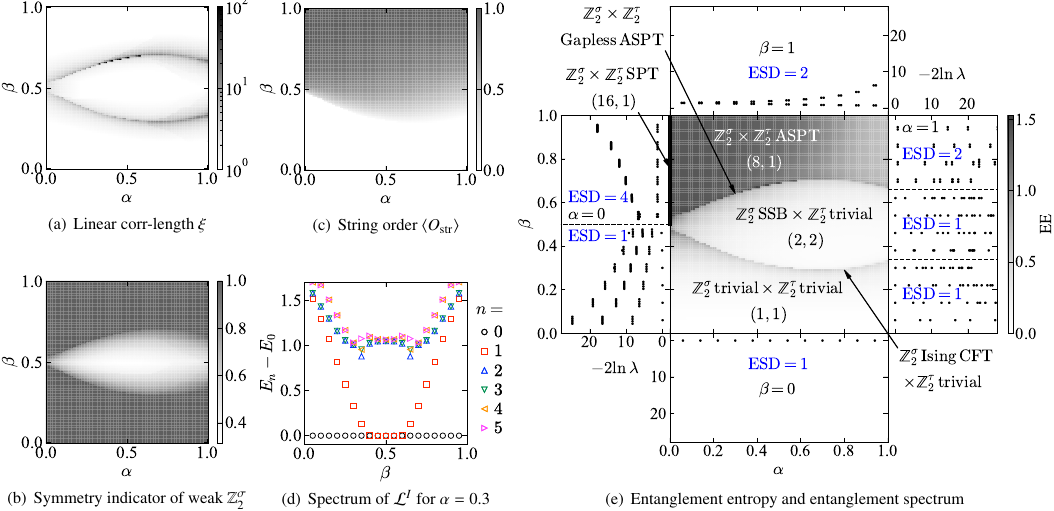}
    \caption{Phase diagram of our model.
    (a) Correlation length $\xi$ of linear correlation functions.
    (b) Symmetry indicator $\braket{U}^{(2)}$ in Eq.~\eqref{Equ: Z2} for $\Z_2^{\sigma}$.
    (c) Expectation value of string order $\braket{O_{\rm str}}$.
    (d) Imaginary-Liouville spectrum for $\alpha=0.3$ and $N=12$ under PBC.
    (e) Entanglement entropy and entanglement spectrum along four axes.
    We mark the ground state degeneracy (GSD) under OBC and entanglement spectrum degeneracy (ESD) for each phase in the diagram.}
    \label{Fig: Phase Diagram}
\end{figure*}

To simulate the steady state resulting from imaginary-time Lindbladian evolution in Eq.~\eqref{Equ: imag}, we employ the time-dependent variational principle (TDVP) method~\cite{Haegeman2011}, which is a cutting-edge technique for calculating time evolution for local Hamiltonians.
The steady-state supervector is approximated using a uniform matrix product state (MPS)~\cite{Perez2007, Orus2014, Haegeman2016, ZaunerStauber2018, Vanderstraeten2019} with physical dimension $d_p^2=16$ and virtual bond dimension $D=16$.
This allows us to label the spins of $\sigma$ and $\tau$ on a single site (Appendix~\ref{App: 1}).

\subsubsection{Correlation lengths}
The divergence of the correlation length is a universal property of quantum criticality in closed systems.
Here, linear and R\'enyi-$2$ correlation lengths (as defined in Eq.~\eqref{Linear_corr} and \eqref{Renyin}, respectively) provide consistent results, as shown in Fig.~\ref{Fig: Phase Diagram}(a) (and Fig.~\ref{Fig: xi2} for $\xi_2$), where two critical lines divide the parameter space into three quantum phases.
The above result suggests that, without rigorous proof at this stage, the divergence of correlation length still serves as an effective indicator for the occurrence of a phase transition in the open system.
Moreover, these two critical lines intersect and end at the point $(\alpha, \beta) = (0, 1/2)$, which is the conventional critical point between the trivial phase and the Haldane phase in the closed system.
In particular, this point can be mapped to the direct product of two $\Z_2$ Ising criticalities using a generalized Kennedy-Tasaki (KT) transformation~\cite{Li2025} under open boundary condition (OBC), i.e.,
\begin{align}
    \mathcal{N}_{\rm KT} H(\alpha=0, \beta)=H^{\prime}(\beta) \mathcal{N}_{\rm KT},
\end{align}
where
\begin{align}
\begin{aligned}
    H^{\prime}(\beta) &= \sum_i \left[(1-\beta) \sigma_i^x+\beta \sigma_i^z\sigma_{i+1}^z\right.\\
    &+ \left.(1-\beta) \tau_{i+1/2}^x+\beta \tau_{i-1/2}^z\tau_{i+1/2}^z\right]\label{Equ: TFIM}
\end{aligned}
\end{align}
denotes two transverse field Ising models (TFIM) with $g=\frac{1-\beta}{\beta}$ and a critical point at $\beta=1/2$.
Therefore, a new phase appears when the system steps into the mixed-state regime, bounded by two critical lines that emerge from this critical point, each of which is expected to belong to the $\Z_2$ Ising universality class.

\subsubsection{Symmetry properties}
To identify the phase of each region, we examine the symmetry properties of the resulting density matrix.
Recall that this system has $\Z_2^{\sigma}\times \Z_2^{\tau}$ symmetry.
Specifically, $K=\prod_i\tau_{i+1/2}^x$ is a strong symmetry satisfying $K\rho = e^{i\theta}\rho$, while $U=\prod_i\sigma_{i}^x$ is a weak symmetry with $U\rho U^{\dagger}=\rho$, or $U\otimes U^*\sket{\rho} = \sket{\rho}$.
To demonstrate whether the density matrix preserves these symmetries, we calculate the following expectations 
\begin{align}
    |\braket{K}| &= \left|\Tr\left[\rho\prod_i\tau_{i+1/2}^x\right]\right|,\\
    \braket{U}^{(2)} &= \frac{\Tr\left[\rho \left(\prod_i{\sigma_i^x}\right) \rho\left(\prod_i{\sigma_i^x}\right)\right]}{\Tr\left[\rho^2\right]}.\label{Equ: Z2}
\end{align}
The choices for these two observables are justified as follows.
If the density matrix preserves the strong symmetry $K\rho=e^{i\theta}\rho$, it is clear that $\braket{K} = e^{i\theta}$.
Conversely, if $|\braket{K}|=1$, we consider the following inequality
\begin{align}
    \left|\braket{K}\right| = \left|\Tr\left[K\sum_i\lambda_i\ket{\psi_i}\hspace{-0.5mm}\bra{\psi_i}\right]\right|\leq \sum_i\lambda_i\left|\braket{\psi_i|K|\psi_i}\right|\leq \sum_i\lambda_i = 1,
\end{align}
where we have used the fact that $K$ is a unitary operator and $\braket{\psi_i|K|\psi_i}$ can be viewed as the overlap between two quantum states.
Therefore, the above equality holds iff $\braket{\psi_i|K|\psi_i}=e^{i\theta}$ for all $\ket{\psi_i}$, which means that $K\rho = e^{i\theta}K$.
For weak symmetry $U$, $\braket{U}^{(2)}$ is equivalent to the inner product between $\rho$ and $U\rho U^{\dagger}$ in the operator space, which equals one iff they are the same density matrix.

Our results show that the first observable $\left|\braket{K}\right|$ always equals one throughout the parameter space, suggesting that the steady state always preserves the $\Z_2^{\tau}$ symmetry.
In contrast, the expectation value of the second observable is depicted in Fig.~\ref{Fig: Phase Diagram}(b), illustrating the SSB of $\Z_2^{\sigma}$ in the intermediate region.
Since there is no nontrivial topological phase solely protected by a single $\Z_2^{\tau}$ (even in the open system~\cite{Ma2023A, Ma2023B, Guo2024A, Xue2024}), the intermediate phase must be $\Z_2^{\sigma}$ SSB $\times$ $\Z_2^{\tau}$ trivial.

\subsubsection{Topological properties}
We now have two regions that preserve both $\Z_2^{\sigma}\times \Z_2^{\tau}$ symmetries.
The upper region includes the points of SPT $\LL^I_{01}$ and ASPT $\LL^I_{11}$, while the lower region includes the points of the trivial pure state $\LL^I_{00}$ and the trivial mixed state $\LL^I_{10}$.
Consequently, we expect that the upper and lower regions represent just the corresponding mixed-state phases of these specific states.

A conventional SPT phase in the closed system exhibits nontrivial topological properties: long-range string order, degenerate entanglement spectrum, and zero energy edge modes under OBC~\cite{Pollmann2010, Pollmann2012}.
These properties are not completely independent of each other and have intrinsic connections.
The former two properties have been discussed in previous work on ASPT~\cite{Ma2023B}, while as for the last one, although the edge mode is related to the degenerate entanglement spectrum, its degeneracy in the energy spectrum has not been well defined using only the state itself without access to a Hamiltonian.
In the following, we will show that this issue can be solved using our imaginary-time Lindbladian formalism, where different edge modes are degenerate regarding the spectrum of $\LL^I$.

We consider the following string order~\cite{Ma2023B}
\begin{align}
    \braket{O_{\rm str}} \equiv \lim_{|i-j|\rightarrow \infty}\braket{\sigma_{i}^z\tau_{i+1/2}^{x}\tau_{i+3/2}^x\cdots \tau_{j-3/2}^x\tau_{j-1/2}^x\sigma_{j}^z},
\end{align}
whose expectation values in different cases (suppose $\Z_2^{\tau} \rm\, symmetric$) are
\begin{align}
    \braket{O_{\rm str}}{\sim}\left\{\begin{array}{ll}
         O(1), & \textrm{domain wall decoration (ASPT)} \\
         \braket{\sigma_i^z\sigma_j^z}, &\textrm{trivial decoration} \sim\left\{\begin{array}{ll}
              O(1), & \Z_2^{\sigma}\rm\, SSB \\
              0, & \Z_2^{\sigma} \rm\, trivial
         \end{array}\right.
    \end{array}\right.
\end{align}
Comparing the results with Fig.~\ref{Fig: Phase Diagram}(c), we conclude that the upper region is an ASPT phase protected by $\Z_2^{\sigma}\times\Z_2^{\tau}$, while the lower region is a trivial phase with the same symmetry.

In addition, we plot the entanglement spectrum (ES) of the supervector $\sket{\rho}$ along the four lines $\alpha=0$, $\alpha=1$, $\beta=0$, and $\beta=1$, together with the entanglement entropy (EE) over the entire parameter space, in Fig.~\ref{Fig: Phase Diagram}(e).
The entanglement spectrum degeneracy (ESD) is labeled with blue marks in each region.
Firstly, the lower panel ($\beta=0$) describes the path connecting the trivial pure state and the trivial mixed state, where all points along the path represent product states with $D=1$.
Secondly, the left panel ($\alpha=0$) shows a conventional phase transition between the trivial phase and the SPT phase (labeled with a thick line) in the closed system, where the ES of the supervector in the SPT phase is four-fold degenerate (square of the ES degeneracy of the corresponding wavefunction).
However, from the upper panel ($\beta=1$) and the right panel ($\alpha=1$), we note that ES is only two-fold degenerate for a general mixed state in the ASPT phase, with the entanglement gap converging to zero when approaching the pure state limit ($\alpha=0$).
Therefore, we deduce that the fundamental property for an ASPT phase protected by $\Z_2^{\sigma} \textrm{(weak)} \times \Z_2^{\tau} \textrm{(strong)}$ is the two-fold degeneracy of the supervector ES, while the four-fold degeneracy in the SPT phase is a nonuniversal property from the additional structure for a closed system and the recovery of $\Z_2^{\sigma}$ strong symmetry, e.g., no coupling between bra and ket.

Finally, we use exact diagonalization (ED) to calculate ground state degeneracy (GSD) under OBC for $N=6$ (i.e., three $\sigma$ spins and three $\tau$ spins), where the results for each phase are marked in Fig.~\ref{Fig: Phase Diagram}(e).
The trivial phase is nondegenerate, and the intermediate SSB phase is two-fold degenerate, both of which match their closed system counterparts.
For the ASPT phase, the two $\sigma$ spins (bra and ket) at the left edge are coupled according to our construction (protected by weak symmetry $U$), while the two $\tau$ spins at the right edge are decoupled (protected by strong symmetry $K$).
This results in three independent spin-$1/2$ to constitute the edge modes and thus $\rm GSD=2^3=8$.
For the pure-state SPT phase ($\alpha=0$), both $\Z_2$ symmetries are strong, and the two $\sigma$ spins are decoupled, leaving us with four independent edge spins and $\rm GSD=2^4=16$.
Similar to the above discussion, we believe that the additional edge mode in the pure-state limit should be interpreted as an accidental degeneracy.
In particular, we emphasize that the edge modes in the ASPT phase are exactly degenerate in a finite-size system, while the degeneracy in the SSB phase only holds in the thermodynamic limit $N\rightarrow \infty$, and a small energy gap exists for a finite $N$.

\subsubsection{Criticaliy and Duality}
The final part discusses the essence of two critical lines in our phase diagram.
The lower critical line describes the transition between $ \rm\Z_2^{\sigma}\,SSB \times \Z_2^{\tau} \,trivial $ and $\rm\Z_2^{\sigma}\, trivial \times \Z_2^{\tau} \, trivial$, corresponding to the conventional Ising criticality.
Meanwhile, the upper critical line between $\rm\Z_2^{\sigma} \times \Z_2^{\tau} \, ASPT$ and $ \rm\Z_2^{\sigma}\, SSB \times \Z_2^{\tau} \, trivial $ is nothing but a mixed-state realization of the gapless SPT proposed in Ref.~\cite{Scaffidi2017} (or more precisely, denoted as gapless ASPT here).
As mentioned above, these two critical lines intersect and end at $(\alpha, \beta) = (0, 1/2)$, the critical point between the SPT phase and the trivial phase.

In addition, in the pure-state case, the gapless SPT can be related to the conventional Ising criticality via a duality map, namely domain-wall (DW) decoration $U_{\rm DW}$, which also converts the trivial product state in Eq.~\eqref{Equ: Trivial} to the cluster state in Eq.~\eqref{Equ: Cluster}~\cite{Scaffidi2017, Li2022, Li2024}.
Our entire phase diagram, especially the positions of two critical lines, shows a symmetry for $\beta\leftrightarrow 1-\beta$, suggesting a similar duality in the general case of mixed states.
To construct the duality for an open system, we need to work on the transformation of the imaginary-Liouville superoperator $\mathcal{L}^I(\alpha, \beta)$.
We still consider the above map $U_{\rm DW}$, which is formally defined as
\begin{align}
    U_{\rm DW} \equiv \prod_{i}\textrm{CZ}_{i-1/2, i}^{\tau, \sigma}\textrm{CZ}_{i, i+1/2}^{\sigma, \tau},
\end{align}
satisfying that
\begin{align}
    U_{\rm DW} \sigma^x_{i} &= \tau^z_{i-1/2}\sigma^x_{i}\tau^z_{i+1/2} U_{\rm DW},\\
    U_{\rm DW} \tau^x_{i+1/2} &= \sigma^z_{i}\tau^x_{i+1/2}\sigma^z_{i+1} U_{\rm DW},\\
    U_{\rm DW} \sigma^z_{i} &= \sigma^z_{i} U_{\rm DW},\\
    U_{\rm DW} \tau^z_{i+1/2} &= \tau^z_{i+1/2} U_{\rm DW}.
\end{align}
From these relations, it is straightforward to verify that
\begin{align}
    U_{\rm DW}H^{10} &= H^{11}U_{\rm DW} ,\\
    U_{\rm DW}L^{11}_{k} &= L^{11}_{k}U_{\rm DW}.
\end{align}
We then derive the duality for our imaginary-Liouville superoperator as
\begin{align}
    U_{\rm DW}\otimes U_{\rm DW}^{*} \LL^I(\alpha, \beta) = \LL^I(\alpha, 1-\beta)U_{\rm DW}\otimes U_{\rm DW}^{*},\label{Equ: DW}
\end{align}
which holds for periodic boundary condition (PBC) or infinite systems.
In particular, $U_{\rm DW}$ is a unitary operator, which means that $\LL^I(\alpha, \beta)$ shares the same spectrum as $\LL^I(\alpha, 1-\beta)$.
For a system without boundaries, both the trivial phase and the ASPT phase are nondegenerate, while the intermediate SSB phase is two-fold degenerate.
Therefore, both critical lines in our phase diagram must be accompanied by the merge of ground states and the first excited states of $\LL^I$ (gap closing) when transiting into the SSB phase, similar to the closed-system case.
The above analysis is verified by the imaginary-Liouville gap $\Delta^I$ depicted in Fig.~\ref{Fig: xi2}(b) calculated for $N=6$ under PBC, where $\Delta^I(\alpha, 1-\beta)=\Delta^I(\alpha, \beta)$ and the intermediate phase possesses a small gap for a finite-size system.
To mitigate the finite-size effect, we perform variational MPS simulation with $D=24$ for a larger system with $N=12$ under PBC and plot the lowest six eigenvalues in Fig.~\ref{Fig: Phase Diagram}(d) along the lines of $\alpha=0.3$ and in Fig.~\ref{Fig: xi2}(c) along the lines of $\alpha=0.6$, respectively.
Both phase transitions in our phase diagram show clear gap-closing patterns, which facilitate the accurate determination of the phase boundary.

In summary, our model simultaneously realizes three typical open-system quantum phases in a single model, together with the criticalities between each other.
Specifically, we identify several topological properties in the ASPT phase, including long-range string order, degenerate entanglement spectrum, and zero-energy edge modes.
We also generalize the duality map between the gapless SPT phase and Ising criticality to the mixed-state formalism, suggesting a systematic method to construct gapless ASPT phases in open systems~\cite{Scaffidi2017, Li2022, Li2024}.

A larger $D=24$ leads to consistent results at several typical points in different phases for all the above calculations.

\section{Comparison between real-time and imaginary-time Lindbladian formalism}
In this section, we continue our analysis of the model introduced previously, with a particular focus on comparing the steady-state properties arising from real-time and imaginary-time Lindbladian evolutions.

\subsection{Steady-state degeneracy}
Firstly, we consider the same set of Hamiltonian in Eq.~\eqref{Equ: Ham} and jump operators in Eq.~\eqref{Equ: Jump} and construct the corresponding real-time Liouville superoperator $\LL$ using Eq.~\eqref{Equ: Liouville}.
In the following, we demonstrate that this real-time formalism does not provide a classification of quantum phases of matter through steady-state properties due to the extensive degeneracy of the steady-state subspace—degeneracy that becomes divergent in the thermodynamic limit $N\rightarrow \infty$.

\begin{figure*}
    \centering
    \includegraphics[width=\linewidth]{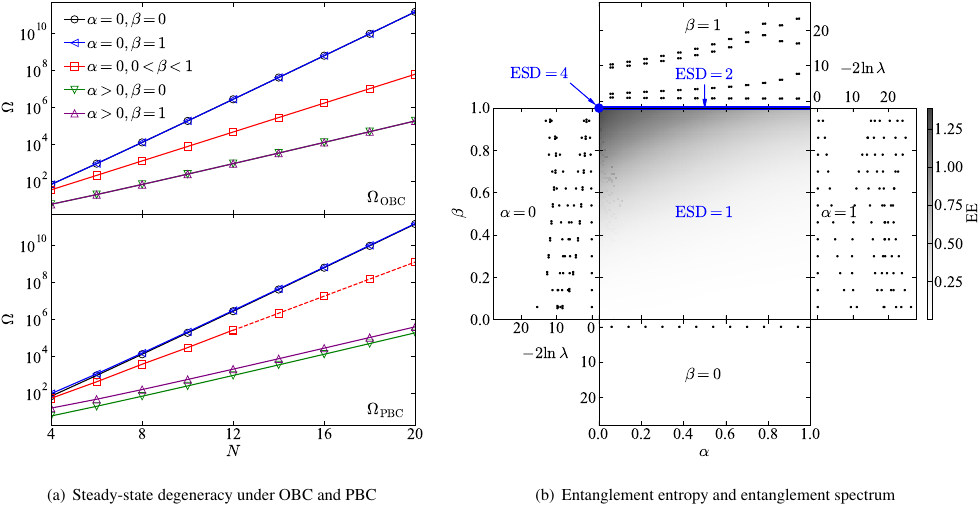}
    \caption{Steady-state properties of real-time imaginary Lindbladian evolution.
    (a) Steady-state degeneracy for the left, upper, and lower panels, using the same set of Hamiltonian in Eq.~\eqref{Equ: Ham} and jump operators in Eq.~\eqref{Equ: Jump}.
    (b) Entanglement entropy and entanglement spectrum along four panels, targeting the same density matrices at four corners.}
    \label{Fig: Real-time}
\end{figure*}

To begin with, we consider the steady-state degeneracy of the left panel ($\alpha=0$), where the real-time Lindbladian evolution reduces to the Schr\"{o}dinger equation and all eigenstates of $H$ are steady states of $e^{\LL t}$.
Specifically, let $E_r$ denote the energy levels of $H$ with degeneracies $\omega_r$, then the steady-state degeneracy of $e^{\LL t}$ reads as
\begin{align}
    \Omega = \sum_r \omega_r^2,
\end{align}
where the total number of eigenstates $\sum_r \omega_r = 2^N$ is just the dimension of Hilbert space.
Therefore, we find that $\Omega\geq 2^N$ for a general closed system with $N$ spins, and this result directly applies to the left panel of our phase diagram.
In particular, the DW duality in Eq.~\eqref{Equ: DW} still holds for the real-time Liouville $\LL$ under PBC, meaning that $\Omega_{\rm PBC}(\alpha, \beta) = \Omega_{\rm PBC}(\alpha, 1-\beta)$.

For the non-interacting point $(\alpha, \beta)=(0, 0)$, the energy levels and the corresponding degeneracies of $N$ spins are
\begin{align}
    E_n = -(N-n) + n= -N+2n,\quad \omega_n = C_N^n,
\end{align}
for $n=0, 1, \cdots, N$ with $\sum_{n=0}^N\omega_n = 2^N$. 
Therefore, the steady-state degeneracy of $e^{\LL t}$ at this point reads as
\begin{align}
    \Omega_{\rm PBC/OBC}(\alpha=0, \beta=0, N) = \sum_{n=0}^N (C_N^n)^2 = C_{2N}^N,
\end{align}
since the boundary condition is irrelevant in the absence of interactions.
Using the DW duality, we immediately reach that
\begin{align}
    \Omega_{\rm PBC}(\alpha=0, \beta=1, N) = \Omega_{\rm PBC}(\alpha=0, \beta=0, N) = C_{2N}^N.
\end{align}

As for the cluster Hamiltonian at $(\alpha, \beta)=(0, 1)$ under OBC, four edge modes provide an extra $16$-fold degeneracy for the final steady-state degeneracy, similar to the previous imaginary-time analysis, while only the bulk $N-2$ degrees of freedom contribute nontrivially to the Hamiltonian, resulting in the following relation
\begin{align}
\begin{aligned}
    &\, \Omega_{\rm OBC}(\alpha=0, \beta=1, N) \\
    =&\, 16\Omega_{\rm PBC}(\alpha=0, \beta=1, N-2) = 16C_{2N-4}^{N-2}
\end{aligned}
\end{align}

Next, we turn to a generic point in the left panel away from the edges, i.e., $\beta\neq 0, 1$.
The spectrum of the Hamiltonian is equivalent to two TFIM in Eq.~\eqref{Equ: TFIM} under OBC, each with $N/2$ spins (when $N$ is even).
Furthermore, each TFIM can be mapped to a free-fermion model via the Jordan-Wigner transformation~\cite{Sachdev2011}, where the single-mode spectrum reads as
\begin{align}
    \varepsilon_k = 2J\sqrt{1+g^2-2g\cos k},\quad g=\frac{1-\beta}{\beta},
\end{align}
and the total energy is expressed as
\begin{align}
    E_{\{n\}} = \sum_{i=1}^{N/2}\varepsilon_{k_i}\left(n_{k_i}^{\sigma}+n_{k_i}^{\tau}-1\right),
\end{align}
where $n_{k_i}^{\sigma (\tau)}=0, 1$ denotes the occupation number of mode $k_i$ associated with $\sigma$ ($\tau$) spins, respectively.
To determine the degeneracy of each many-body energy level, we note that a single mode $k_i$ has three energy levels $E^{k_i}_{r_i}$ with degeneracy $\omega^{k_i}_{r_i}$ listed as follows (we have take the modes from $\sigma$ and $\tau$ spins as a whole)
\begin{align}
\begin{aligned}
    &E^{k_i}_{0} = -\varepsilon_{k_i}, &\omega_0^{k_i} = 1,\\
    &E^{k_i}_{1} = 0, &\omega_1^{k_i} = 2,\\
    &E^{k_i}_{2} = \varepsilon_{k_i}, &\omega_2^{k_i} = 1.
\end{aligned}
\end{align}
Therefore, for a given configuration ${\{r\}}=(r_1, \cdots, r_{N/2})$ of mode energies, the total degeneracy is
\begin{align}
    \omega_{\{r\}}=\prod_{i=1}^{N/2}\omega^{k_i}_{r_i}.
\end{align}
We obtain the steady-state degeneracy of $e^{\LL t}$
\begin{align}
\begin{aligned}
    &\,\Omega_{\rm OBC}(\alpha=0, 0<\beta<1, N)=\sum_{\{r\}}\omega_{\{r\}}^2\\
    =&\,\sum_{\{r\}} \prod_{i=1}^{N/2}(\omega^{k_i}_{r_i})^2=\prod_{i=1}^{N/2}\left[\sum_{r_i} (\omega^{k_i}_{r_i})^2\right] = 6^{N/2}.
\end{aligned}
\end{align}

Finally, we tune on the dissipative terms and examine the mixed steady states of a general real-time Lindbladian evolution.
The system remains non-interacting along the lower panel $(\beta=0, \alpha>0)$, where dissipative terms lead to a unique steady state for $\sigma$ spins, while $\tau$ spins evolve unitarily as a closed system.
Consequently, the total steady-state subspace is still highly degenerate.
Specifically, we have
\begin{align}    
\begin{aligned}
    &\, \Omega_{\rm PBC/OBC}(\alpha>0, \beta=0, N) \\
    =&\, \Omega_{\rm PBC/OBC}(\alpha=0, \beta=0, N/2) = C_{N}^{N/2},
\end{aligned}
\end{align}
from which we also obtain the degeneracy along the upper panel under PBC using the DW duality
\begin{align}
    \Omega_{\rm PBC}(\alpha>0, \beta=1, N) = \Omega_{\rm PBC}(\alpha>0, \beta=0, N) = C_{N}^{N/2}.
\end{align}
As for the upper panel under OBC, three edge modes contribute an additional factor of $2^3=8$ to the steady-state degeneracy.
This arises from two independent $\tau$ spins and one $\sigma$ degree of freedom (two $\sigma$ spins in the ket and bra are coupled) for a general mixed state.
The remaining $(N/2-1)$ $\tau$ spins contribute according to their closed-system evolution.
Thus, the total steady-state degeneracy becomes
\begin{align}
\begin{aligned}
    &\, \Omega_{\rm OBC}(\alpha>0, \beta=1, N) \\
    =&\, 8\Omega_{\rm PBC}(\alpha>0, \beta=1, N-2) = 8C_{N-2}^{N/2-1}
\end{aligned}
\end{align}
The above results are summarized in the top and bottom panels of Fig.~\ref{Fig: Real-time}(a) corresponding to OBC and PBC, respectively.
Here, an analytical expression for $\Omega_{\rm PBC}$ ($\alpha=0$, $0<\beta<1$, $N$) is difficult to obtain. 
Therefore, we plot the numerical results with ED for $N\leq12$ and then linearly extrapolate to larger system sizes (dashed lines).
Notably, all of them scale exponentially with the system size $N$, corresponding to a highly degenerate steady-state manifold.

\subsection{Real-time phase diagram}
The above section demonstrates that if starting from the same set of $H$ and $L_k$, the real-time Lindbladian typically possesses a highly degenerate steady-state manifold even when entering the mixed-state regime, hindering any reasonable identification of possible phase transitions.
An alternative approach to comparing the imaginary-time and real-time frameworks is to consider Lindbladians that share the same steady-state density matrix, while allowing for different choices of $H$ and $L_k$.
To this end, we adopt a procedure analogous to that outlined around Eqs.~\eqref{Equ: HL1}-\eqref{Equ: Jump}.

The real-time $H$ and $L_k$ to realize the four corner mixed states can be constructed as follows~\cite{Wang2024}
\begin{align}
    \textrm{Trivial pure: } & L_i^{[1]00}=\sigma_{i}^z\frac{1-\sigma_{i}^x}{2}\sigma_{i+1}^z,\nonumber\\
    & L_i^{[2]00}=\tau_{i-1/2}^z\frac{1-\tau_{i-1/2}^x}{2}\tau_{i+1/2}^z,\\
    \textrm{SPT: } & L_i^{[1]01}=\sigma_{i}^z\frac{1-\tau_{i-1/2}^z\sigma_{i}^x\tau_{i+1/2}^z}{2}\sigma_{i+1}^z,\nonumber\\
    & L_i^{[2]01}=\tau_{i-1/2}^z\frac{1-\sigma_{i-1}^z\tau_{i-1/2}^x\sigma_i^z}{2}\tau_{i+1/2}^z,\\
    \textrm{Trivial mixed: } & L^{[1]10}_{i} = \sigma_i^z, \ L^{[2]10}_{i} = \sigma_i^x,\nonumber\\
    & L_i^{[3]10}=\tau_{i-1/2}^z\frac{1-\tau_{i-1/2}^x}{2}\tau_{i+1/2}^z\\
    \textrm{ASPT: } & L^{[1]11}_{i} = \sigma_i^z, \ L^{[2]11}_{i} = \tau_{i-1/2}^z\sigma_i^x\tau_{i+1/2}^z,\nonumber\\
    & L_i^{[3]11}=\tau_{i-1/2}^z\frac{1-\sigma_{i-1}^z\tau_{i-1/2}^x\sigma_i^z}{2}\tau_{i+1/2}^z,
\end{align}
with all Hamiltonians being zero for the above four corners.
Subsequently, we construct the real-time Liouville superoperators $\LL$ according to Eq.~\eqref{Equ: Liouville} and apply a similar bilinear interpolation
\begin{align}
\begin{aligned}
    \LL(\alpha, \beta) &= (1-\alpha)(1-\beta)\LL_{00}+(1-\alpha)\beta \LL_{01} \\
    & +\alpha(1-\beta)\LL_{10}+\alpha\beta\LL_{11}.
\end{aligned}
\end{align}

The steady states for different parameter values are computed using the infinite time-evolving block decimation (iTEBD) method, whose EE are depicted in the phase diagram shown in Fig.~\ref{Fig: Real-time}(b), which displays a smooth and continuous distribution across the entire parameter space, with no indication of singular behavior.
Moreover, we plot the ES along four panels and identify ESD for different regimes.
The results clearly show that nontrivial ASPT order—signaled by a characteristic two-fold degeneracy in the ES—appears only along the upper panel of the diagram.
In contrast, all other regions correspond to trivial symmetric phases.
Notably, as in the imaginary-time phase diagram shown in Fig.~\ref{Fig: Phase Diagram}(e), the ES at the top-left corner exhibits a four-fold degeneracy.
This feature stems from the restoration of strong $\Z_2^{\sigma}$ symmetry in the pure-state limit, which is not a universal property for the ASPT phase focused on here.
In conclusion, these results demonstrate that the real-time Lindbladian formalism fails to capture the essential phase transitions between trivial and ASPT phases.
This is because the corresponding steady states can be continuously deformed into one another within the real-time phase diagram, obscuring any topological distinction.

\section{Conclusions and Discussions}
In this paper, we propose a new framework to understand quantum phases and phase transitions in an open system governed by the Lindbladian equation.
We introduce the concept of imaginary-time Lindbladian evolution to connect the mixed state and the system described by an imaginary-Liouville superoperator $\LL^I$, completing the last piece of the puzzle for open-system quantum phases.
Our example not only demonstrates several important topological phenomena for ASPT phases but also sheds light on some universal properties in a general phase transition, such as nonanalytic physical observables, divergence of correlation lengths, and closing of the imaginary-Liouville gap.
In contrast, the steady states of real-time Lindbladian evolution sometimes do not directly reflect the phase distinctions captured by the Markov length or imaginary-time framework, which is essential to the current understanding of mixed-state topological phases.
Nevertheless, real-time Lindbladians remain indispensable in the study of realistic open quantum systems, as they describe the dissipative dynamics resulting from system-bath couplings.
We believe that our results can complement this conventional approach and open a new avenue for systematic research on quantum phase transitions in open systems.

On the other hand, we emphasize that our construction of the phase diagram follows a bottom-up approach.
Namely, we begin by explicitly constructing states of interest and then investigate phase transitions between distinct topological phases by interpolating the corresponding Liouville superoperators.
This strategy aligns with the widely used parent Hamiltonian method, which has proven effective in the classification and construction of topological phases in both closed systems and non-Hermitian settings~\cite{Perez2007, Shen2023}.
Alternatively, a more conventional top-down approach can be employed, wherein one starts from a standard closed-system Hamiltonian and introduces suitable jump operators to model system-environment coupling, typically with a tunable dissipation strength.
Within the imaginary-time Lindbladian framework, this allows for the systematic construction of phase diagrams starting from well-known Hamiltonians.
As a concrete example, we present this construction in Appendix~\ref{App: 2}, where we analyze a dissipative TFIM with jump operators $L_i=\sqrt{\alpha}\sigma_i^y$.
This example illustrates how Ising criticality manifests and evolves upon entering the mixed-state regime.

Several avenues emerge for further study.
Firstly, we have explicitly constructed the imaginary-Liouville $\LL^I$ to describe a finite-temperature Gibbs state for a given stabilizer Hamiltonian.
The construction of a more general case, if possible, will further extend the application scope of our formalism to characterize a general mixed state.
Secondly, we demonstrate the coincidence between the closing of the imaginary-Liouville gap and the occurrence of phase transition in 1D systems with Hermitian jump operators $L_k$.
As shown in our 1D numerical example, phase transitions are consistently accompanied by the divergence of two types of correlation length $\xi$ and $\xi_2$ that are numerically accessible, thus validating their role as effective indicators for mixed-state criticality.
However, extending this result to higher spatial dimensions remains a significant challenge.
At the heart of this issue lies the relationship among different orders of R\'enyi correlation lengths $\xi_n$, where we believe that at least one of them diverges at any phase transition in an open system.
A rigorous proof of this argument or a counterexample (a phase transition occurs where all orders of R'enyi correlation lengths remain finite) will be of great significance for a better understanding of the essence of open-system phase transitions.

\begin{acknowledgments}
    We thank Hao-Ran Zhang, Jian-Hao Zhang, and Zhong-Xia Shang for helpful discussions. This work is supported by the National Natural Science Foundation of China (NSFC) (Grant No. 12475022 and No. 12174214) and the Innovation Program for Quantum Science and Technology (Grant No. 2021ZD0302100).
\end{acknowledgments}

\bibliography{ref}
\newpage
\appendix
\onecolumngrid
\renewcommand{\theequation}{S\arabic{equation}} \setcounter{equation}{0}
\renewcommand{\thefigure}{S\arabic{figure}} \setcounter{figure}{0}
\section{Two types of correlation function and more results for the phase diagram}\label{App: 1}

In this section, we will illustrate the concrete steps to evaluate two types of expectation values in Eq.~\eqref{Linear_corr} and \eqref{Renyin}, which are calculated on a density matrix.

\begin{figure}[H]
    \centering
    \includegraphics[width=0.6\linewidth]{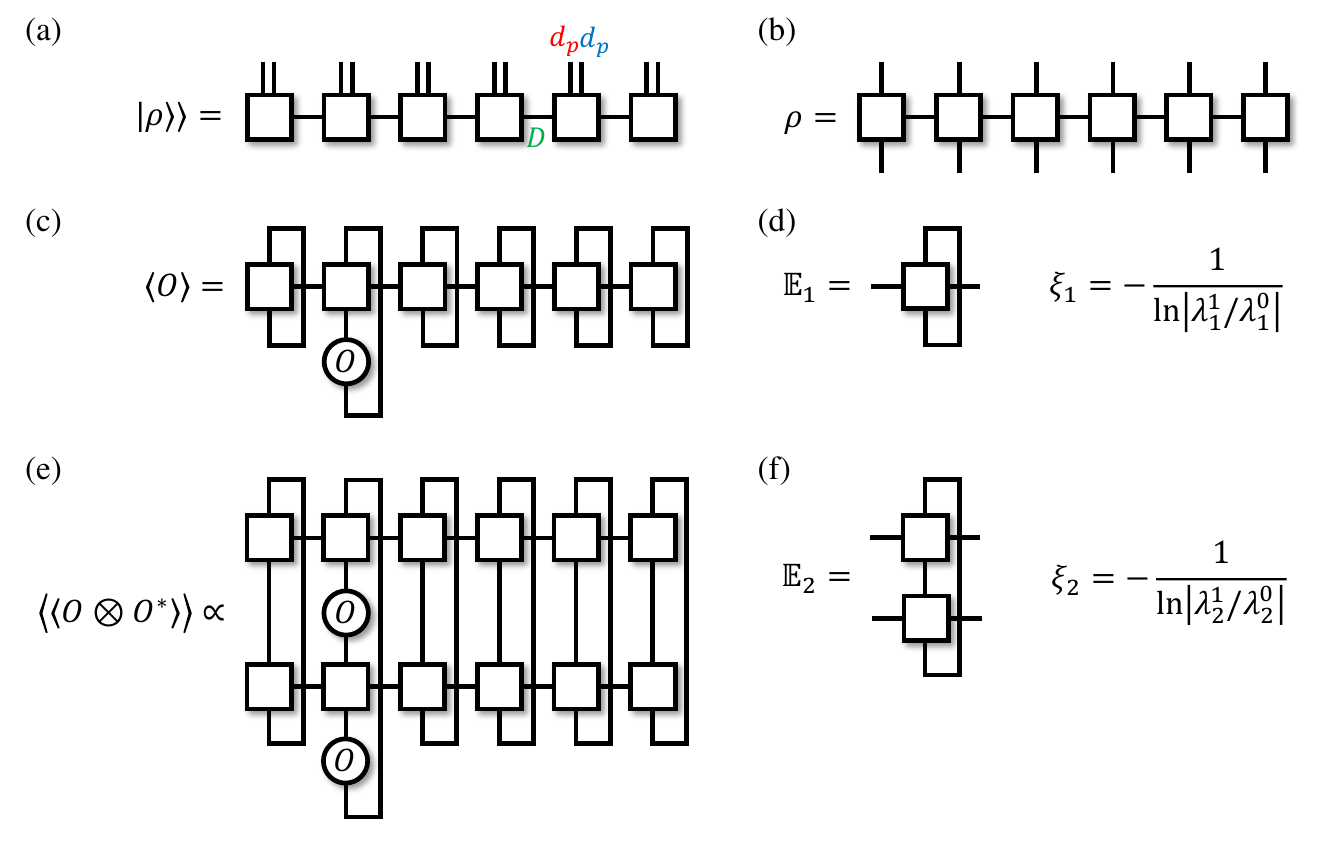}
    \caption{Calculation of correlation lengths.
    (a) The uniform MPS to represent the supervector $\sket{\rho}$.
    (b) The MPO to represent the density matrix $\rho$.
    (c) The linear expectation values $\braket{O}$.
    (d) The linear transfer matrix $\mathbb{E}$ and the correlation length $\xi$.
    (e) The R\'enyi-2 expectation values $\braket{O}^{(2)}$.
    (f) The R\'enyi-2 transfer matrix $\mathbb{E}_2$ and correlation length $\xi_2$.}
    \label{Fig: App}
\end{figure}

\begin{figure}[H]
    \centering
    \includegraphics[width=0.75\linewidth]{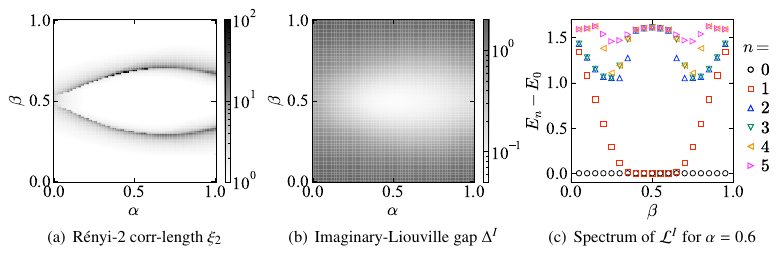}
    \caption{More results for the phase diagram in Fig.~\ref{Fig: Phase Diagram}.
    (a) Correlation length $\xi_2$ of R\'enyi-2 correlation functions for our model.
    (b) Imaginary-Liouville gap $\Delta^I$ for $N=6$ under PBC.
    (c) Imaginary-Liouville spectrum for $\alpha=0.6$ and $N=12$ under PBC.}
    \label{Fig: xi2}
\end{figure}

Firstly, we consider a uniform MPS shown in Fig.~\ref{Fig: App}(a) with two physical indices (each with $d_p=4$ labeling $\sigma$ and $\tau$ spins on a single site) to represent the ket and bra of a density matrix.
Once we obtain the MPS representation for the supervector $\sket{\rho}$ of the steady state of our imaginary-time Lindbladian evolution $e^{-\int \LL^I\mathrm{d}\tau}$, we divide the physical indices for the ket and bra and obtain the matrix product operator (MPO) representation for a density matrix in Fig.~\ref{Fig: App}(b).
In this way, both observables in Eq.~\eqref{Linear_corr} and \eqref{Renyin}  can be efficiently calculated with the standard tensor contraction technique, as illustrated in Fig.~\ref{Fig: App}(c) and (e).
Moreover, the correlation lengths for the correlators defined in the above two schemes $\xi$ and $\xi_2$ are determined by the transfer matrices $\mathbb{E}_1$ and $\mathbb{E}_2$ defined in Fig.~\ref{Fig: App}(d) and (f), respectively.
Specifically, $\lambda_i^0$ and $\lambda_i^1$ are the eigenvalues of $\mathbb{E}_i$ with the largest and second largest magnitude for $i=1,\ 2$.
The R\'enyi-2 correlation length $\xi_2$ for our model is shown in Fig.~\ref{Fig: xi2}, which exhibits a similar divergent trend as the standard linear correlation length $\xi$ shown in Fig.~\ref{Fig: Phase Diagram}(a).
Meanwhile, we plot the imaginary-Liouville gap $\Delta^I$ calculated using ED for $N=6$ under PBC in Fig.~\ref{Fig: xi2}(b), as well as the spectrum of $\LL^I$ along the line of $\alpha=0.6$ (the lowest six eigenvalues) calculated using variational MPS with $D=24$ for $N=12$ under PBC.
These results provide a clear illustration of the key features in our phase diagrams, including the degeneracy in the SSB phase as well as the divergence of correlation length and the closing of the imaginary-Liouville gap along the critical lines.

\section{Dissipative Ising model}\label{App: 2}
\begin{figure}
    \centering
    \includegraphics[width=0.5\linewidth]{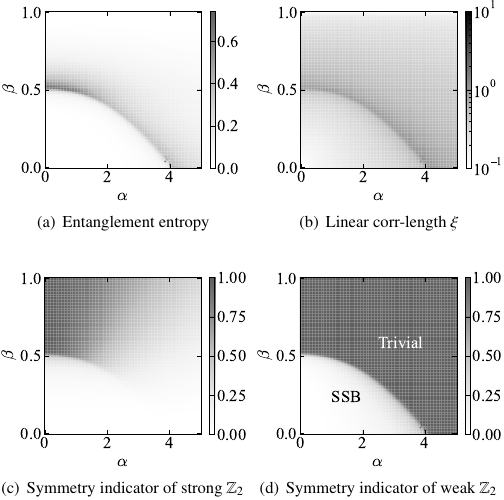}
    \caption{Phase diagram for dissipative TFIM.
    (a) Correlation length $\xi$ of linear correlation functions.
    (b) Entanglement entropy.
    (c) Symmetry indicator $|\braket{U}|$ for strong $\Z_2$ symmetry.
    (d) Symmetry indicator $\braket{U}^{(2)}$ for weak $\Z_2$ symmetry.}
    \label{Fig: TFIM}
\end{figure}

Here, we calculate the imaginary-time Lindbladian phase diagram for the dissipative TFIM, whose Hamiltonian and jump operator are given by
\begin{align}
    H(\beta) &= -\beta\sum_i\sigma_i^{x} - (1-\beta)\sum_i\sigma_i^z\sigma_{i+1}^z,\\
    L_i &= \sqrt{\alpha}\sigma_i^y,
\end{align}
from which we construct the imaginary-Liouville superoperator $\LL^I(\alpha, \beta)$.
This open system preserves a weak $\Z_2$ symmetry generated by $U=\prod_i \sigma_i^x$, whose phase diagram is shown in Fig.~\ref{Fig: TFIM}.
The point $(\alpha, \beta)=(0, 0.5)$ corresponds to the conventional phase transition between a trivial symmetric phase and an SSB phase.
As the dissipation strength $\alpha$ increases, this critical point extends to the mixed-state regime, forming a critical line that intersects the lower panel ($\beta=0$) near $(\alpha, \beta)\sim (4, 0)$.
This critical line is identified by the divergence of both EE and $\xi$ in Fig.~\ref{Fig: TFIM}(a-b).
Meanwhile, the symmetry indicators for strong and weak $\Z_2$ symmetries presented in Fig.~\ref{Fig: TFIM} reveal that the introduction of the dissipative term explicitly breaks the strong $\Z_2$ symmetry into a weak one.
The aforementioned critical line then separates the phase diagram into a trivial (weakly) symmetric phase and an SSB phase that further breaks the weak symmetry.
This example illustrates how the imaginary-time Lindbladian framework captures phase transitions in open systems in a way that is consistent with conventional closed-system phase transitions by adding a parameterized dissipative term.
\end{document}